\documentclass[%
 aip,
rsi,%
 amsmath,amssymb,
 reprint,%
superscriptaddress
]{revtex4-1}

\usepackage{graphicx}
\usepackage{dcolumn}
\usepackage{bm}
\usepackage{hyperref}
\usepackage{caption}
\usepackage{color}

\usepackage{color}

\graphicspath{{Figures/}{./}}

\begin{document}


\title[]{Chimera states in multi-strain epidemic models with temporary immunity}

\author{Larissa Bauer}
\email{larissa.c.bauer@googlemail.com}
\affiliation{Institut f{\"u}r Theoretische Physik, Technische Universit{\"a}t 
Berlin, Hardenbergstra\ss{}e 36, 10623 Berlin, Germany
}%

\author{Jason Bassett}%
\email{j.bassett@tu-berlin.de}
\affiliation{Institut f{\"u}r Theoretische Physik, Technische Universit{\"a}t 
Berlin, Hardenbergstra\ss{}e 36, 10623 Berlin, Germany
}%
\author{Philipp H\"{o}vel} 
\email{phoevel@physik.tu-berlin.de}
\affiliation{Institut f{\"u}r Theoretische Physik, Technische Universit{\"a}t 
Berlin, Hardenbergstra\ss{}e 36, 10623 Berlin, Germany
}%
\affiliation{Bernstein Center for Computational Neuroscience Berlin,
Humboldt-Universit\"{a}t zu Berlin, Philippstra\ss e 13, 10115 Berlin, Germany}

\author{Yuliya N. Kyrychko} 
\email{y.kyrychko@sussex.ac.uk}
\affiliation{%
Department of Mathematics, University of Sussex, Falmer, Brighton BN1 9QH, United Kingdom
}%

\author{Konstantin B. Blyuss}
\email{k.blyuss@sussex.ac.uk}
\affiliation{%
Department of Mathematics, University of Sussex, Falmer, Brighton BN1 9QH, United Kingdom
}%

\date{\today}

\begin{abstract}
We investigate a time-delayed epidemic model for multi-strain diseases with temporary immunity. In the absence of cross-immunity between strains, dynamics of each individual strain exhibits emergence and annihilation of limit cycles due to a Hopf bifurcation of the endemic equilibrium, and a saddle-node bifurcation of limit cycles depending on the time delay associated with duration of temporary immunity. Effects of all-to-all and non-local coupling topologies are systematically investigated by means of numerical simulations, and they suggest that cross-immunity is able to induce a diverse range of complex dynamical behaviors and synchronization patterns, including discrete traveling waves, solitary states, and amplitude chimeras. Interestingly, chimera states are observed for narrower cross-immunity kernels, which can have profound implications for understanding the dynamics of multi-strain diseases.
\end{abstract}

\pacs{05.45.Xt, 05.45.-a, 87.23.Cc, 89.75.-k}
\keywords{multi-strain epidemic model, cross-immunity, coupled oscillators, non-local coupling, chimera states}
\maketitle

\begin{quotation}
One of the most fascinating phenomena that has intrigued researchers in the area of nonlinear dynamics for the last fifteen years is a very peculiar pattern of behavior known as {\it chimera states}, which is characterized by the simultaneous coexistence of regions of coherent and incoherent dynamics. This pattern was found when identical oscillators were connected with a non-local coupling of high symmetry. In the following years chimera states have attracted a lot of interest and have been studied theoretically and experimentally in a variety of different contexts. This paper investigates how chimera states can appear in epidemic models, and it also explores wider dynamics of multi-strain diseases with time delay and non-local coupling. 
\end{quotation}

\section{Introduction}
\label{sec:back}

Chimera is a hybrid state with coherent and incoherent dynamics, which was first described by Kuramoto and Battogtokh in a system of coupled identical oscillators \cite{KUR02a}. This unusual dynamical pattern was called a {\it chimera state} by Abrams and Strogatz \cite{ABR04} in light of analogy with a mythological creature with three heads of three different animals. Chimera states have been subsequently discovered in various contexts: SQUID materials \cite{LAZ15}, quantum systems \cite{BAS15}, electronic oscillators \cite{GAM14}, and many more. It is currently debated that the dynamics observed, for instance, in uni-hemispheric sleep in mammals and birds \cite{RAT00}, and blackouts in power-grids \cite{FIL08a,MAR13,MOT13a} can be interpreted as chimera states.
Whilst chimera states have been observed in a number of natural phenomena, they are quite complicated to implement experimentally for several reasons. Firstly, only small networks can be realized in laboratory conditions, and identical oscillators with identical intrinsic frequencies are required \cite{OME15}. Secondly, chimera states can be very sensitive to initial conditions and often occur only in a small region of the parameter space, and, thus, an experimental setup has to be very precisely controlled in terms of all parameters. Recent studies on two coupled populations of phase oscillators have also demonstrated the possibility of extended basins of attraction \cite{MAR16c}, and the existence of chimeras even for small numbers of elements \cite{PAN16c}. Thirdly, chimera states are a transient state that collapses after a finite period of time into a state of full synchrony \cite{WOL11}. Although the lifetime of chimeras has been reported to increase exponentially \cite{WOL11} or as a power-law \cite{OLM15a,OLM15} in dependence on the number of oscillators, it can be very short for small networks. Despite these challenges, chimera states have been robustly produced in several experiments, including chemical oscillators \cite{NKO13}, optical systems \cite{HAG12}, time-delayed laser networks \cite{LAR15}, electrochemical oscillators \cite{LAR13}, and mechanical oscillators \cite{MAR13}. For a recent review, see Panaggio and Abrams \cite{PAN15}.

The formation and properties of chimera states have been studied in a number of theoretical models represented as networks of FitzHugh-Nagumo \cite{OME13}, Kuramoto \cite{OLM15a, WAN11f}, Ginzburg-Landau \cite{SET13}, van der Pol \cite{ULO16}, leaky integrate-and-fire \cite{TSI16}, Stuart-Landau \cite{ZAK15b}, Hindmarsh-Rose \cite{HIZ13}, Hodgkin-Huxley \cite{GLA16}, and SNIPER \cite{VUE14} oscillators, as well as many other models. Whilst originally chimera states were discovered in the case of non-local coupling \cite{KUR02a}, subsequently a number of other topologies have been identified that can result in chimera states, including global \cite{BOE15} and local \cite{LAI15,BER16a} coupling. There is a large variety in manifestations of chimera states and how they can appear in different systems. If one considers amplitude-phase representation of individual node dynamics, it is possible to distinguish between phase chimeras and amplitude chimeras. The phase chimera is defined as the coexistence of coherent and incoherent regions in the space of phases of different oscillators \cite{BOG16a}. In this case, the average phase velocity of different oscillators exhibits a characteristic arc-shape profile, with a pronounced increase or decrease in the average frequency for the incoherent region associated with the chimera state. In contrast, an amplitude chimera appears as a sudden increase or decrease in the average amplitude of oscillations \cite{SET13, ZAK14, BOG16a}. 

In this paper, we consider the emergence and behavior of chimeras in the specific context of epidemic models of multi-strain diseases. A number of effective mathematical frameworks have been developed over the years for the analysis of various aspects of strain interactions \cite{AND97,GUP98,GOG02,GOG02a,GOM02,KOE06,KOE09}, with particular attention being paid to cross-immunity and its effects \cite{CAL05,ADA07,MIN08a,COB11}. Multi-strain epidemic models have been shown to exhibit a wide range of behaviors, including (partially) synchronized dynamics, anti-phase oscillations, as well as chaotic dynamics \cite{MIN08a,REC09,KUC16}. Group-theoretical analysis of multi-strain models has yielded significant inroads to systematic classification of steady states and periodic solutions in terms of their symmetry \cite{BLY12,BLY13,BLY14,CHA13a,CHA13b}. Motivated by the recent work on chimeras in locally coupled, delayed oscillators \cite{BER16a}, we explore the dynamics of a multi-strain network, in which coupling between strains quantifies the degree of their cross-immunity, while the dynamics of each individual strain is represented by a compartmental model, with the time delay representing a period of temporary immunity upon recovery from infection.


The remainder of this paper is organized as follows. In the next Section we introduce the model and discuss its basic properties. Section ~\ref{sec:uncoupled} contains analytical and numerical bifurcation studies of single-strain dynamics for completely antigenically distinct strains. In Section~\ref{sec:coupled} different types of dynamics are investigated in the presence of all-to-all and non-local cross-immunity coupling kernels. The paper concludes in Section~\ref{sec:conclusion} with the discussion of results.

\section{Model}
\label{sec:model}


We consider a multi-strain disease, in which recovery from an infection with any single strain results in a certain period of temporary immunity against subsequent infections with that strain. To analyze the dynamics of such a disease, one can combine an SIRS-type model of temporary immunity proposed by Kyrychko and Blyuss \cite{KYR05,BLY10} with the status-based approach of Gog and Grenfell \cite{GOG02} for multi-strain diseases, which gives the following model
\begin{equation}
\label{eq:SIR}
\begin{array}{l}
	\dot{S}_{i}(t) =\eta - \eta S_{i}(t)-S_{i}(t)\sum\limits_{j=1}^{N}\beta_{j}\sigma_{ij}I_{j}(t)+\gamma_i I_{i}(t-\tau) e ^{-\eta \tau},\\
	\dot{I}_{i}(t) =\beta_{i}S_{i}(t)I_{i}(t)-(\gamma_i + \eta) I_{i}(t),\\\\
	\dot{R}_{i}(t) =\gamma_i I_{i}(t) - \gamma_i I_{i}(t-\tau)e^{- \eta \tau} - \eta R_{i}(t)\\\\
	\hspace{1.3cm}+ S_{i}(t) \sum\limits_{j=1, j\neq i}^{N} \beta_{j}\sigma_{ij}I_{j}(t),\\
\end{array}
\end{equation}
where $S_i$, $I_i$ and $R_i$ represent the number of people in the population that are susceptible, infected or recovered from strain $i=1, 2, \dots, N$, with $N$ being the total number of disease strains in circulation, $\eta > 0$ is a constant birth rate and death rate assumed to be the same for all strains, $\beta_{i} > 0$ and $\gamma_i> 0$  are the transmission rate and the recovery rate of strain $i$, respectively. This model assumes that after recovery, individuals remain the class of recovered from strain $i$ for a period of temporary immunity $\tau > 0$, upon which they return to the class of susceptible. For simplicity, we assume that the transmission and recovery rates for all strains are the same, namely, $\beta_{i}=\beta$ and $\gamma_i=\gamma$. The factor $0 \leq \sigma_{ij} \leq 1$ denotes the reduction in the susceptibility to strain $i$ due to immune response to a previous infection with strain $j$ \cite{GOM02}, with zero denoting the complete cross-immunity, that is, the same immunological response between two strains $i$ and $j$, and unity denoting the complete absence of cross-immunity, i.e., absolutely distinct immunological responses against the two strains $i$ and $j$. In this paper, we will consider all-to-all coupling, i.e., $\sigma_{ij}\equiv 1$, as well as two types of non-local coupling kernels that represent more realistic immunological relations between disease strains.

Summation of the left- and right-hand sides of Eqs.~\eqref{eq:SIR} yields
\begin{align}
  \dot{S}_i(t)+\dot{I}_i(t)+\dot{R}_i(t) &= \eta - \eta \left[S_i(t)+I_i(t)+R_i(t)\right]\\
 \Leftrightarrow \quad \dot{N}_i(t) &= \eta \left[1 -N_i(t)\right],
\end{align}
where $N_i(t)=S_i(t)+I_i(t)+R_i(t)$ denotes the total population of strain $i$.
Since the birth and death rates are equal, the total population for each strain is asymptotically constant \cite{GOG02,KYR05}, that is, all $N_i$ tend to unity.
The observations that $R_i(t)=1-S_i(t)-I_i(t)$ and that $R_i(t)$ does not feature in equations for $S_i$ and $I_i$, suggest that it is sufficient to focus on the dynamics of variables $S_{i}$ and $I_{i}$ only. To reduce the number of free parameters, we rescale time with $(\eta + \gamma)^{-1}$, and introduce a basic reproduction number $r=\beta/(\eta+\gamma)$ and a rescaled mortality rate $\rho=\eta/(\eta + \gamma)$. This gives the following rescaled model
\begin{equation}\label{eq:SI}
\begin{array}{l}
 \dot{S}_{i}(t) =\rho [1-S_{i}(t)]-rS_{i}(t)I_{i}(t) + (1-\rho)I_{i}(t-\tau )e^{-\rho \tau}\\\\ 
 \hspace{1.3cm}-rS_{i}(t)\sum\limits_{j=1, j\neq i}^{N} \sigma_{ij}I_{j}(t),\\\\
 \dot{I}_{i}(t) = rS_{i}(t)I_{i}(t)-I_{i}(t),
\end{array}
\end{equation}
where the self-coupling term is written out explicitly with $\sigma_{ii}=1$.

\section{Antigenically distinct strains}
\label{sec:uncoupled}

Before investigating the collective behavior in the full multi-strain system, it is instructive to consider what happens
in the absence of cross-immunity, i.e., when each strain is genetically distinct, so as to cause a completely distinct immunological response to infection, which is represented by $\sigma_{ij}=\delta_{ij}$, where $\delta_{ij}$ is the Kronecker delta. In this case, the system~\eqref{eq:SI} decouples into $N$ independent copies, and the dynamics of each individual strain is described by the following system of equations
\begin{equation}
\label{eq:SI_single}
\begin{array}{l}
	\dot{S}(t) = \rho [1-S(t)]-rS(t)I(t) + (1-\rho)I(t-\tau )e^{-\rho \tau},\\\\
	\dot{I}(t) = rS(t)I(t)-I(t).
\end{array}
\end{equation}
This system always has the disease-free steady state $E_0=(S_{0},I_{0})=(1,0)$, and it can also possess an endemic steady state
\begin{equation}\label{endemic_fxd_pnt}
E^*_{\tau}=(S^{*}_{\tau},I^{*}_{\tau}) = \left(\frac{1}{r},\rho \frac{r-1}{r}\frac{1}{1-(1-\rho)e^{-\rho \tau}}\right).
\end{equation}
The endemic equilibrium $E^*_{\tau}$ is only biologically feasible if $r>1$, which, in terms of original parameters, corresponds to the transmission rate $\beta$ being larger than the sum of the natural death rate $\eta$ and the recovery rate $\gamma$.
In the case of very long immunity period, i.e. for $\tau \rightarrow \infty$, the SIRS model~\eqref{eq:SI_single} transforms into a standard SIR model with vital dynamics and permanent immunity, and the 
endemic steady state then reduces to 
\begin{equation}\label{SI_end_state}
E^{*}_{\infty} = 
\left(S^{*}_{\infty},I^{*}_{\infty}\right) = \left(\frac{1}{r},\rho \frac{r-1}{r}\right).
\end{equation}



Linearization of the system~\eqref{eq:SI_single} near the disease-free steady state $E_0$ gives the characteristic eigenvalues as $\lambda_{1} = -\rho$ and $\lambda_{2} = r-1$, thus implying that the disease-free steady state is stable, provided $r<1$. For the endemic steady state $E^*_{\tau}$, the characteristic equation has the form 
\begin{equation}\label{endem_charc_eq_uncoup}
\lambda^{2} + \lambda(\rho + rI^{*}_{\tau}) - [rI^{*}_{\tau}-\rho (r-1)e^{-\lambda \tau}]=0,
\end{equation}
which, for a vanishing delay $\tau=0$, always gives stable eigenvalues due to $r>1$. One root of this equation is $\lambda=-\rho<0$, which is stable independently of the time delay. For non-zero immunity period, the endemic steady state can lose its stability in a Hopf bifurcation, giving rise to periodic solutions.

Since for $\tau=0$ the eigenvalues $\lambda$ of the characteristic equation~\eqref{endem_charc_eq_uncoup} are stable, and $\lambda=0$ is never a solution of this equation, the only possibility how the stability of the endemic steady state can change is if a pair of complex conjugate eigenvalues crosses the imaginary axis for some value of $\tau$. To find this critical time delay, we substitute $\lambda = i \omega$ into Eq.~\eqref{endem_charc_eq_uncoup}
and separate real and imaginary parts, which yields
\begin{align}\label{omega_Re_Im_system}
	-\omega ^{2} + r I^{*}_{\tau} = [rI^{*}_{\tau} -\rho (r-1) \cos (\omega \tau)],\nonumber \\
	\omega (\rho +rI^{*}_{\tau}) = [rI^{*}_{\tau} + \rho (r-1) \sin (\omega \tau)].
\end{align}
Squaring and adding these two equations gives an implicit equation for the Hopf frequency
\begin{align}
	&\omega^{4} + \omega^{2} [\rho^{2} -2rI^{*}_{\tau} (1-\rho) + r^{2} (I^{*}_{\tau})^{2}] \nonumber \\
	& -\rho (r-1) [\rho (r-1) -2rI^{*}_{\tau}]=0,
\end{align}
which can be readily solved to give
\begin{widetext}
\begin{align}\label{eq:omega}
	\omega_{\pm}^{2}  = \frac{1}{2} \left[ -\rho^{2} +2rI^{*}_{\tau} (1-\rho) - r^{2} (I^{*}_{\tau})^{2} \right]  
	  \pm \sqrt{\left[\rho^{2} -2rI^{*}_{\tau} (1-\rho) + r^{2} (I^{*}_{\tau})^{2}\right]^{2} 
	 + 4\rho (r-1) [\rho (r-1) -2rI^{*}_{\tau}]}.
\end{align}
\end{widetext}

Alternatively, by dividing the equations~\eqref{omega_Re_Im_system}, we find the critical value of the time delay at which the Hopf bifurcation occurs
\begin{equation}
  \tau_{c} = \frac{1}{\omega}\left[ \arctan \left(\frac{\omega(\rho + r I^{*}_{\tau_c})}{\omega ^{2} - rI^{*}_{\tau_c}} \right) + n\pi \right], \hspace{0.3cm}n \in \mathbb{N}.
\end{equation}
Unfortunately, due to the fact that the steady-state value of the infected fraction $I^*_{\tau}$ itself explicitly depends on the time delay $\tau$ as shown in Eq.~\eqref{endemic_fxd_pnt}, it does not prove possible to find a closed form expression for the Hopf frequency or the critical time delay. 

\begin{figure}[ht!]
	\includegraphics[width=\linewidth]{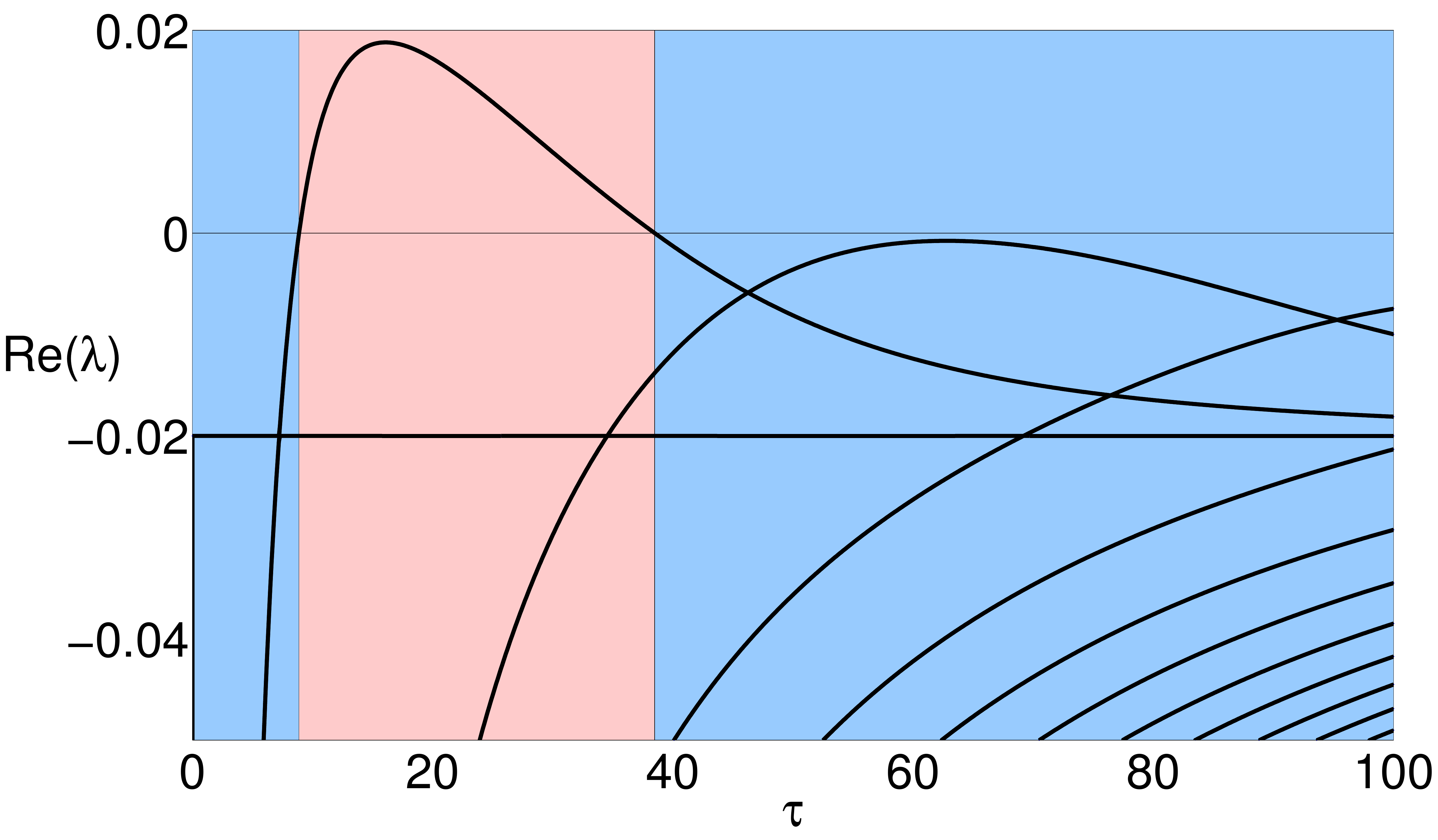}
	\caption{\label{fig:re_lamb_vs_tau} Real part of the eigenvalue versus time delay. The blue and red areas indicate regions of stability and instability of the endemic steady state, respectively. Parameter values are $\rho=0.02$ and $r=2$.}
\end{figure}

To get a better understanding of the bifurcations of the endemic fixed point, we perform numerical bifurcation continuation using \textsc{DDE-Biftool} \cite{ENG01}, choosing $\tau$ as the continuation parameter. Figure~\ref{fig:re_lamb_vs_tau} illustrates regions of stability and instability of this steady states, together with multiple branches of characteristic eigenvalues.
For the chosen parameter values, this figure shows that a single branch escapes the stable region from $\tau_{1} \approx 8.88$ to $\tau_{2} \approx 38.49$, and in this interval of time delays, the endemic steady state is unstable.

\begin{figure}[ht!]
	\includegraphics[width=\linewidth]{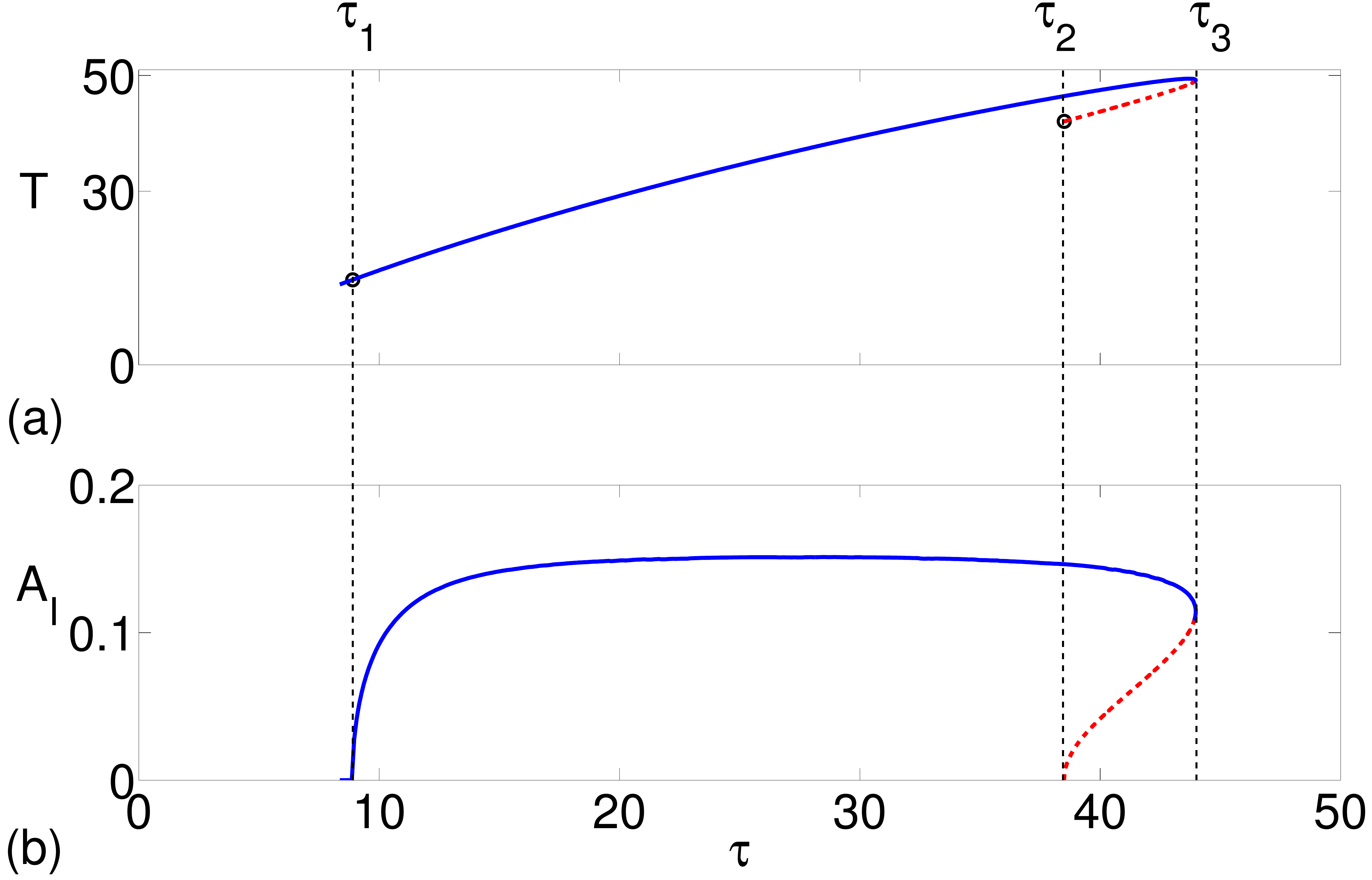}
	\caption{\label{fig:lim_cycle_figs} Period $T$ and amplitude $A_I$ of the limit cycle as a function of the delay $\tau$ are shown in panels (a) and (b), respectively. Blue (solid) and red (dashed) lines correspond to stable and unstable limit cycles, respectively. The vertical lines mark critical time delays $\tau_{1}$ and $\tau_{2}$ for the super- and sub-critical Hopf bifurcation, as well as $\tau_{3}$, at which a saddle-node bifurcation of limit cycles occurs. Parameters as in Fig.~\ref{fig:re_lamb_vs_tau}.}
\end{figure}

Having identified the points at which the endemic equilibrium loses/gains its stability, we now focus on the limit cycle that emerges at these bifurcation points.
Figure~\ref{fig:lim_cycle_figs} shows the period and amplitude of the limit cycle depending on the time delay. This figure indicates that at $\tau=\tau_1\approx 8.88$, the endemic steady state undergoes a supercritical Hopf bifurcation, giving rise to a stable limit cycle, whereas at $\tau=\tau_{2} \approx 38.49$ it undergoes a subcritical Hopf bifurcation, at which the endemic equilibrium regains its stability, and an unstable limit cycle is born. These two limit cycles coexist for $\tau>\tau_{2}$ until they merge at a point $\tau_{3} \approx 43.97$ and annihilate in a saddle-node bifurcation of limit cycles.




\section{Multi-strain dynamics}\label{multiple_strain_sims}
\label{sec:coupled}

As a next step, we consider the network of $N$ coupled strains~\eqref{eq:SI}, where in the absence of coupling the dynamics of each strain is described by a delayed SIR model~\eqref{eq:SI_single}. Before proceeding with numerical simulations, it is worth noting that for any form of the coupling $\sigma_{ij}$, the system~\eqref{eq:SI} admits a one-strain solution with $I_{i}(t)\neq 0$ and $I_{j}(t)=0$ for $j\neq i$ that defines an invariant manifold (cf. Blyuss \& Gupta \cite{BLY09} for a similar type of behavior in a $\mathbb{D}_4$-symmetric model of antigenic variation), and whose behavior is described by the following system
\begin{equation}\label{one_strain_dyn}
\begin{array}{l}
 \dot{S}_{i}(t) =\rho [1-S_{i}(t)]-rS_{i}(t)I_{i}(t) + (1-\rho)I_{i}(t-\tau )e^{-\rho \tau},\\\\
 \dot{I}_{i}(t) =rS_{i}(t)I_{i}(t)-I_{i}(t),\\\\
 \dot{S}_{j}(t) =\rho [1-S_{j}(t)]-r\sigma_{ji}S_{j}(t)I_{i}(t),\hspace{0.3cm}j\neq i.\\
\end{array}
\end{equation}
Effectively, the system decouples into the single-strain dynamics~\eqref{eq:SI_single} for strain $i$, which then drives the evolution of $S_j$ variables, while all $I_j$ remain zero. The equivalent one-strain endemic steady state is given by
\begin{equation}\label{one_strain_ss}
\begin{array}{l}
\displaystyle{S_i^{*}=\frac{1}{r},\hspace{0.3cm}I_i^{*}=\rho \frac{r-1}{r}\frac{1}{1-(1-\rho)e^{-\rho \tau}},}\\\\
\displaystyle{S_j^{*}=\frac{1-(1-\rho)e^{-\rho \tau}}{1-(1-\rho)e^{-\rho \tau}+\sigma_{ij}(r-1)},\hspace{0.3cm}I_j^{*}=0,\hspace{0.3cm}j\neq i.}
\end{array}
\end{equation}
In the case of all-to-all coupling with $\sigma_{ij}=1$, the system~\eqref{eq:SI} possesses a $\mathbb{Z}_N$ symmetry, hence it has $N$ identical one-strain steady states given by Eq.~\eqref{one_strain_ss} for any $i=1,...,N$. Furthermore, for such coupling the system~\eqref{one_strain_dyn} reduces to just strain $i$ with the dynamics given by Eq.~\eqref{eq:SI_single}, and all other strains, whose dynamics is exactly the same and is fully driven by the strain $i$. Techniques of equivariant bifurcation theory can be used to systematically characterize various steady states and periodic solutions in terms of their symmetry \cite{BLY12,BLY13,BLY09,BLY14}.

Besides one-strain steady states, the system~\eqref{eq:SI} also has a fully symmetric endemic steady state
\begin{equation}
S_1^{*}=...=S^{*}_N=S_{\rm end}^{*},\hspace{0.3cm}I_1^{*}=...=I_N^{*}=I_{\rm end}^{*},
\end{equation}
where
\[
S_{\rm end}^{*}=\frac{1}{r},\hspace{0.3cm}I_{\rm end}^{*}=\rho \frac{r-1}{r}\frac{1}{1-(1-\rho)e^{-\rho \tau}+\sigma_{c}},
\]
with $\sigma_{c}=\sum\limits_{i\ne j}\sigma_{ij}$.

For each type of coupling, we have used the \textsc{dde23} solver\cite{SHA01a} to numerically integrate the system~\eqref{eq:SI} with the initial conditions taken as follows: $S_i$ are uniformly distributed random numbers between $0$ and $1$ independent for each strain, and random $I_i\in[0,1-S_i]$ being constant in $t\in[\tau,0)$. We investigate possible dynamical behavior for three different types of coupling between strains: the all-to-all coupling, a Gaussian kernel based on the model of Gog and Grenfell \cite{GOG02}, and a functional cosine kernel suggested by Gomes et al. \cite{GOM02} Since the last two kernels are non-local, in principle, one can expect to observe chimera states in such multi-strain systems \cite{KUR02a, ABR04}, and below we investigate the appearance of such states and transitions between them and other dynamical regimes.


\subsection{All-to-all coupling}

In the case of global all-to-all coupling $\sigma_{ij}=1$, the same amount of cross-immunity is present between all interacting strains, which biologically means that every strain is related to all other strains in the same way.
\begin{figure}[ht!]
	\includegraphics[width=\linewidth]{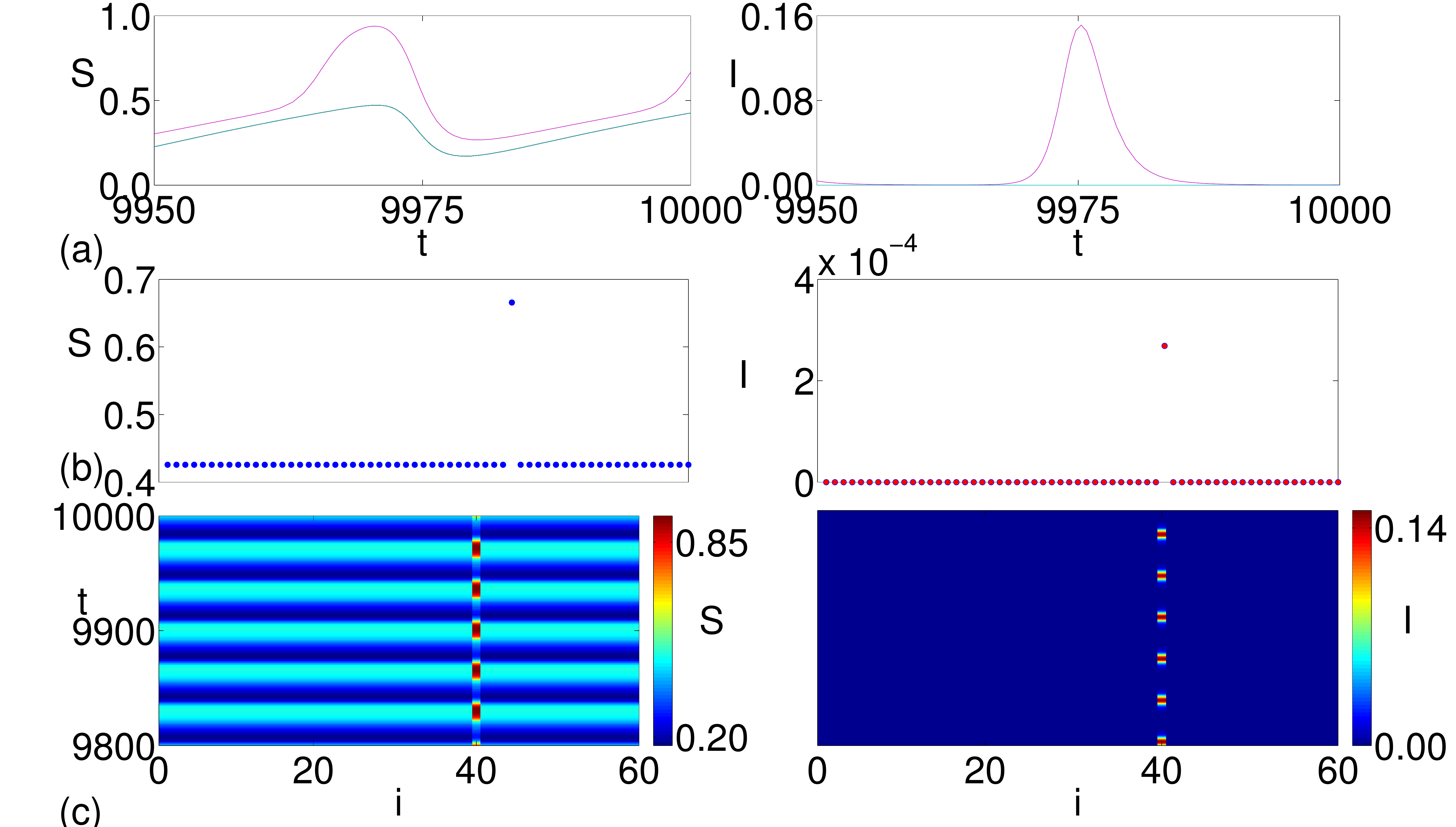}
	\caption{\label{fig:all-to-all} All-to-all coupling, i.e., $\sigma_{ij} = 1$, with the time delay $\tau = 25$. Panels (a) show the time series of $S_i$ and $I_i$ for $N=60$ strains, where the solitary strain (here: $i=40$) is shown in purple. Panels (b) illustrate snapshots at a fixed time, and panels (c) are the respective space-time plots. Other parameters as in Fig.~\ref{fig:re_lamb_vs_tau}.}
\end{figure}
Figure~\ref{fig:all-to-all} shows the dynamics of system~\eqref{eq:SI} with $N=60$ strains for an all-to-all coupling and time delay $\tau = 25$, for which a stable limit cycle is observed in the single-strain dynamics. The time series, as well as the snapshot and the space-time plot, indicate that in this case all nodes become synchronized, except for one strain ($i=40$ here), as shown in Fig.~\ref{fig:all-to-all}. The latter strain exhibits large-amplitude oscillations in both $S$ and $I$ variables, which then drive smaller amplitude oscillations in the $S$ variable for all other strains.
As discussed earlier, the dynamics of such a {\it solitary state} can be effectively described by a reduced two-strain model: one delayed model~\eqref{eq:SI_single} for the solitary strain, and one for all other synchronized strains, as given in Eq.~\eqref{one_strain_dyn}. 

One should note that due to the above-mentioned $\mathbb{Z}_N$ symmetry of the system, the fact that the system has settled on the strain $i=40$ being the main driving strain is completely random and is purely determined by the initial conditions, as for the same parameter values, any of the other solitary states is equally possible. The other observation is that since the system starts with random and independent initial conditions for all strains, the fact that eventually it settles on a solitary state suggests that a one-strain invariant manifold described by Eq.~\eqref{one_strain_dyn} is stable. Moreover, since this  corresponds to a situation where in the absence of coupling all individual strains have the dynamics of a stable limit cycle, effectively the coupling appears to suppress these oscillations in a manner similar to symmetry-breaking oscillations death that has been recently studied in time-delayed systems \cite{ZAK13}.


\subsection{Non-local kernels}

\begin{figure}[ht!]
	\includegraphics[width=\linewidth]{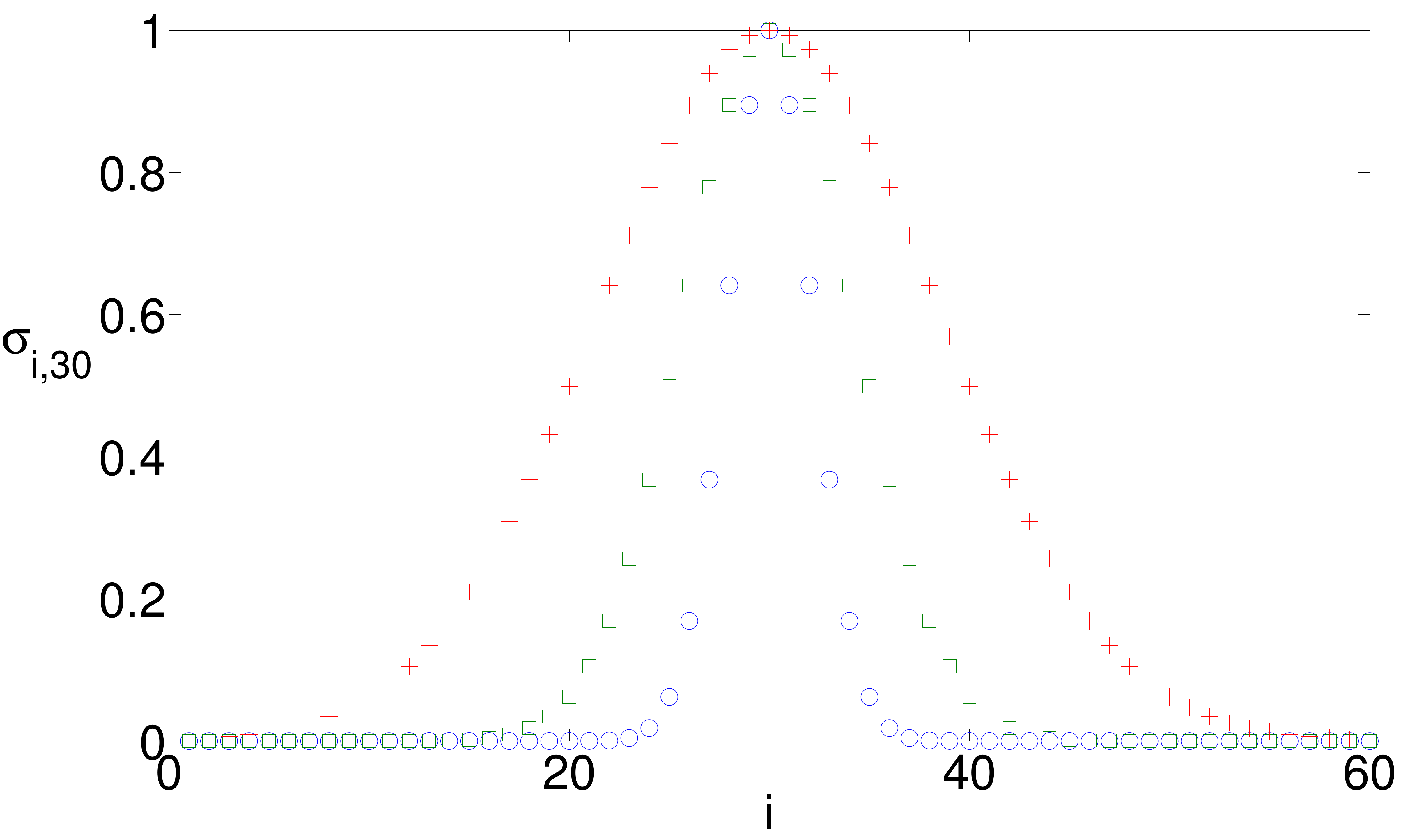}
	\caption{\label{fig:Gaus_coup} Gaussian coupling kernel $\sigma_{ij}$ described by Eq.~\eqref{eq:Gauss_sigma} for $N=60$, with respect to strain $j=30$, and three different values of the characteristic length $d=3$ (blue circles), $6$ (green squares), and $12$ (red crosses).}
\end{figure}

By analogy with non-local coupling kernels for which chimera states have been observed in various systems of coupled oscillators \cite{KUR02a,ABR04,OME10a}, we focus our attention on two kernels that represent the biologically realistic scenario where the more related strains are, the higher is the level of cross-immunity between them \cite{GOG02,GOM02}. The first example is a slightly modified Gaussian kernel introduced in Gog \& Grenfell~\cite{GOG02}
\begin{equation}\label{eq:Gauss_sigma}
    \sigma_{ij}= \exp \left(-\frac{\left[\frac{N}{2}-\min (|j-i|,N-|j-i|)\right]^{2}}{d^{2}}\right), 
\end{equation}
where $d$ is the characteristic length associated with cross-immunity, and the distance between strains $i$ and $j$ is measured as the smallest difference on the interval $[1,N]$ with periodic boundary conditions.
Strains that are genetically close to each other have a higher value of cross-immunity $1-\sigma_{ij}$, leading to a decrease in the inflow of the infected population for the strain at hand. This effect is a combination of the reduced susceptibility and reduced infectivity due to various immunological interactions between strains \cite{GOG02,BAL09a}. Figure~\ref{fig:Gaus_coup} illustrates the shape of the kernel $\sigma_{ij}$ for different characteristic lengths $d$.

\begin{figure}[ht!]
	\includegraphics[width=\linewidth]{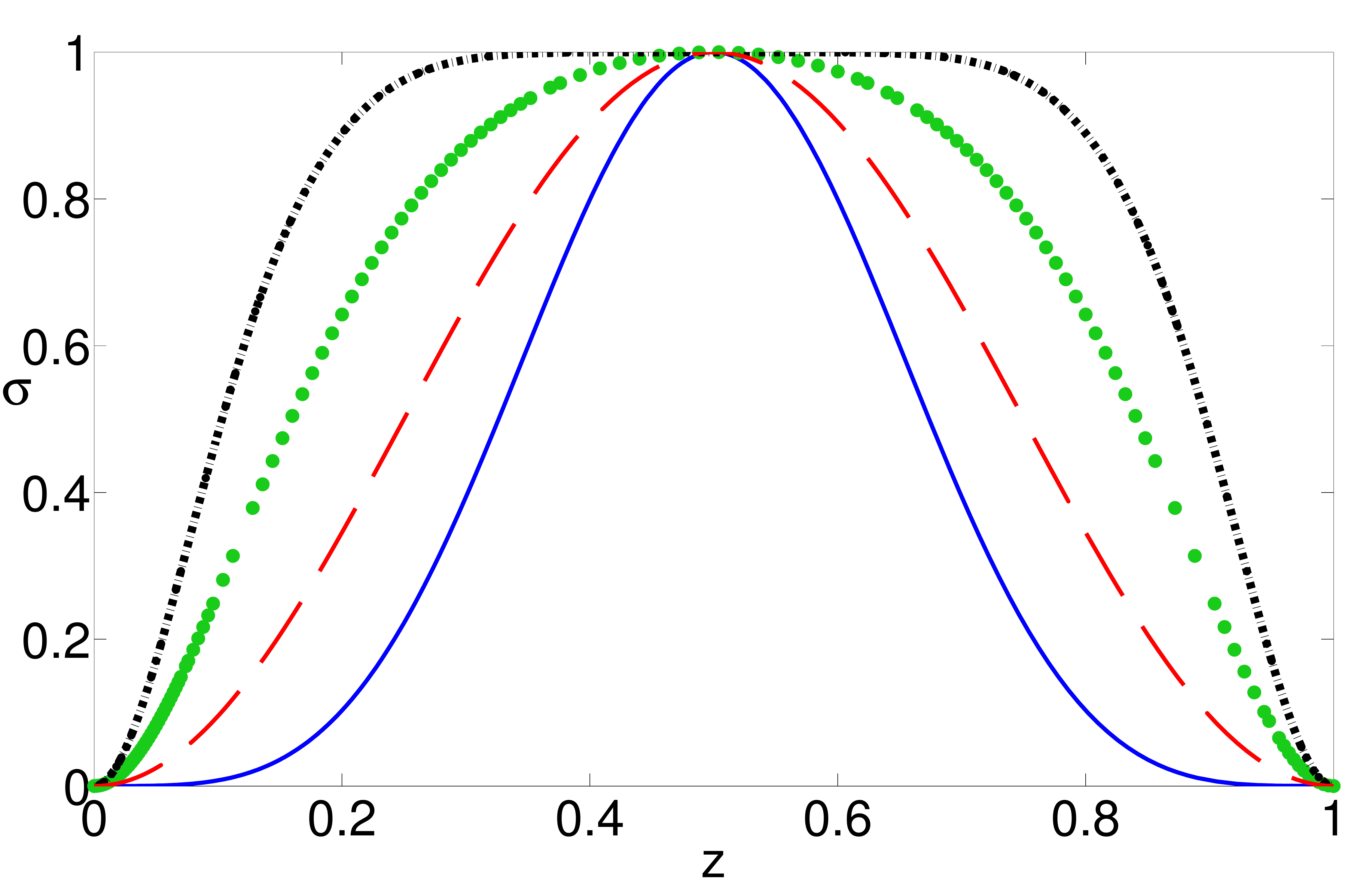}
	\caption{\label{fig:cos_func} Cosine coupling kernel $\sigma$ satisfying the continuous form of Eqs.~\eqref{eq:cos_coup},\eqref{eq:cos_coup_dp} for different values of $p$: $p=-2$ (blue solid), $0$ (red dashed), $2$ (green dotted), and $4$ (black dash-dotted).}
\end{figure}

A similar reasoning, but with a different biological rationale, is used in the model of Gomes et al.~\cite{GOM02} who considered strains as being distributed on the unit circle with positions $z_{i}=(2i-1)/2N$ along the circle, with the kernel being given by
\begin{equation}\label{eq:cos_coup}
    \sigma_{ij}=\frac{\sigma}{2}\left\{1-\cos \left[ 2\pi d_{p} \left( \min \left[ |z_{j}-z_{i}|,z_{N}-|z_{j}-z_{i}| \right]  \right)  \right] \right\}
\end{equation}
with
\begin{equation}\label{eq:cos_coup_dp}
   d_{p}(z)=z+pz\left(z-\frac{1}{2}\right)(z-1).
\end{equation}
The profile of $\sigma_{ij}$ depending on the distance between strains is illustrated in Fig.~\ref{fig:cos_func} for different values of parameter $p$.

In the coupling kernel~\eqref{eq:cos_coup} there are two different parameters that characterize the strain space. Firstly, there is $\sigma$ ($0 \leq \sigma \leq 1$), which plays the role of the bound on the range of the strain diversity. Secondly, there is $p$ which represents antigenic differences between strains for the given genetic range. Gomes et al. \cite{GOM02} focused on the specific values of $p=-2$, $0$ and $2$, but one can prove that parameter $p$ must lie in the range $p \in [-2,4]$ to ensure $\sigma(z)$ has a single maximum at $z=0.5$ and two minima at $z=0$ and $z=1$, which biologically means that the strain most genetically different from the current strain experiences the smallest amount of cross-immunity. 

\subsection{Emergent dynamical scenarios}

\begin{figure}[ht!]
	\includegraphics[width=\linewidth]{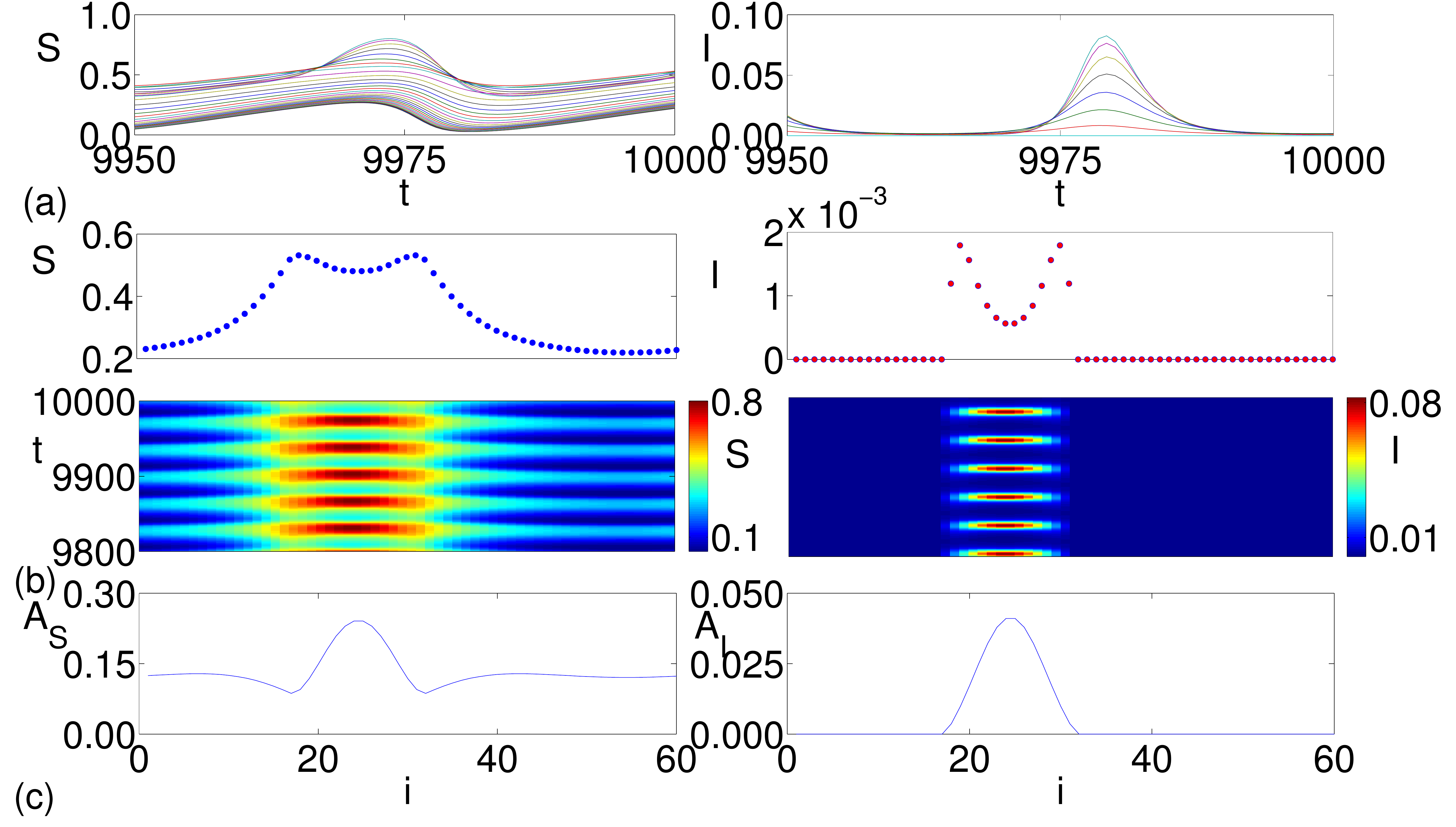}
	\caption{\label{fig:gauss_modulated} Modulated profile: time series, snapshots, space-time plots, and amplitude profiles for system~\eqref{eq:SI} with Gaussian coupling given by Eq.~\eqref{eq:Gauss_sigma}. Coupling parameters are $\tau = 25, d = 14$, and $\sigma = 0.7$, with other parameters as in Fig.~\ref{fig:re_lamb_vs_tau}.}
\end{figure}

\begin{figure}[ht!]
	\includegraphics[width=\linewidth]{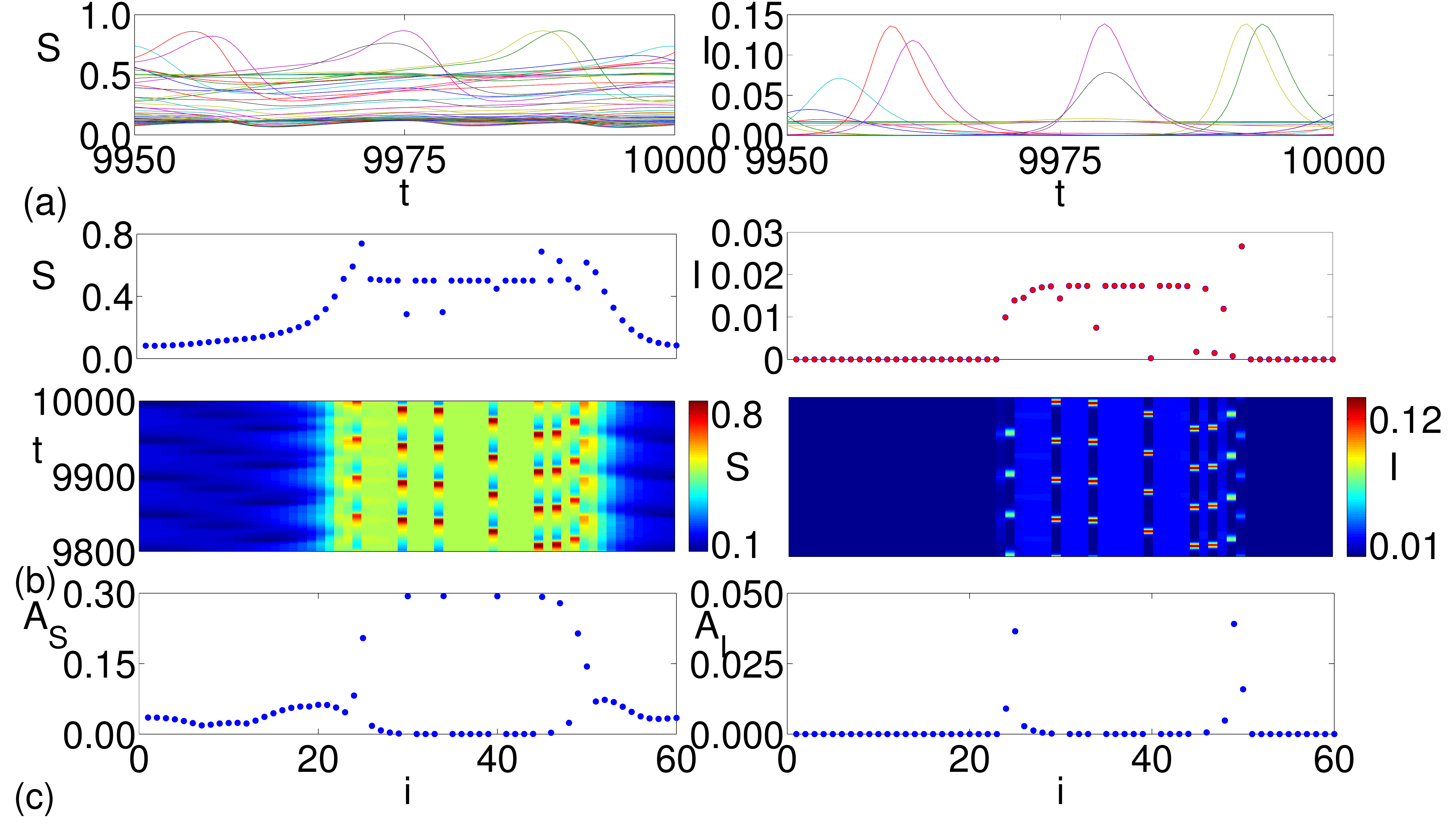}
	\caption{\label{fig:gauss_solitary} Solitary states: time series, snapshots, space-time plots, and amplitude profiles for system~\eqref{eq:SI} with Gaussian coupling given by Eq.~\eqref{eq:Gauss_sigma}. Coupling parameters are $\tau = 42, d = 4$, and $\sigma = 0.7$, with other parameters as in Fig.~\ref{fig:re_lamb_vs_tau}. }
\end{figure}

\begin{figure}[ht!]
	\includegraphics[width=\linewidth]{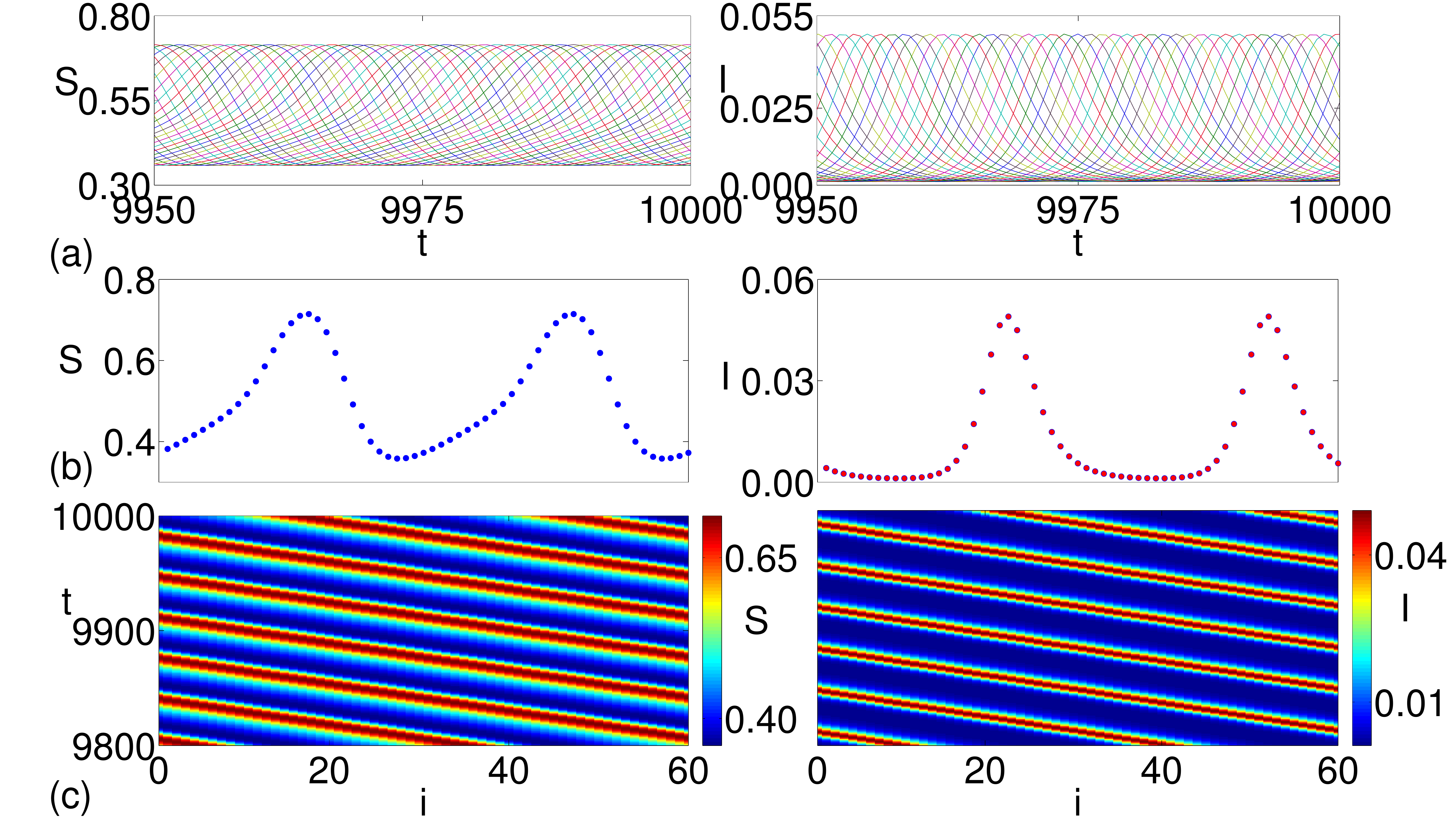}
	\caption{\label{fig:gauss_wave} Traveling wave: time series, snapshots, and space-time plots for system~\eqref{eq:SI} with Gaussian coupling given by Eq.~\eqref{eq:Gauss_sigma}. Coupling parameters: $\tau = 25, d = 2$, and $\sigma = 0.1$, with other parameters as in Fig.~\ref{fig:re_lamb_vs_tau}.}
\end{figure}

\begin{figure}[ht!]
	\includegraphics[width=\linewidth]{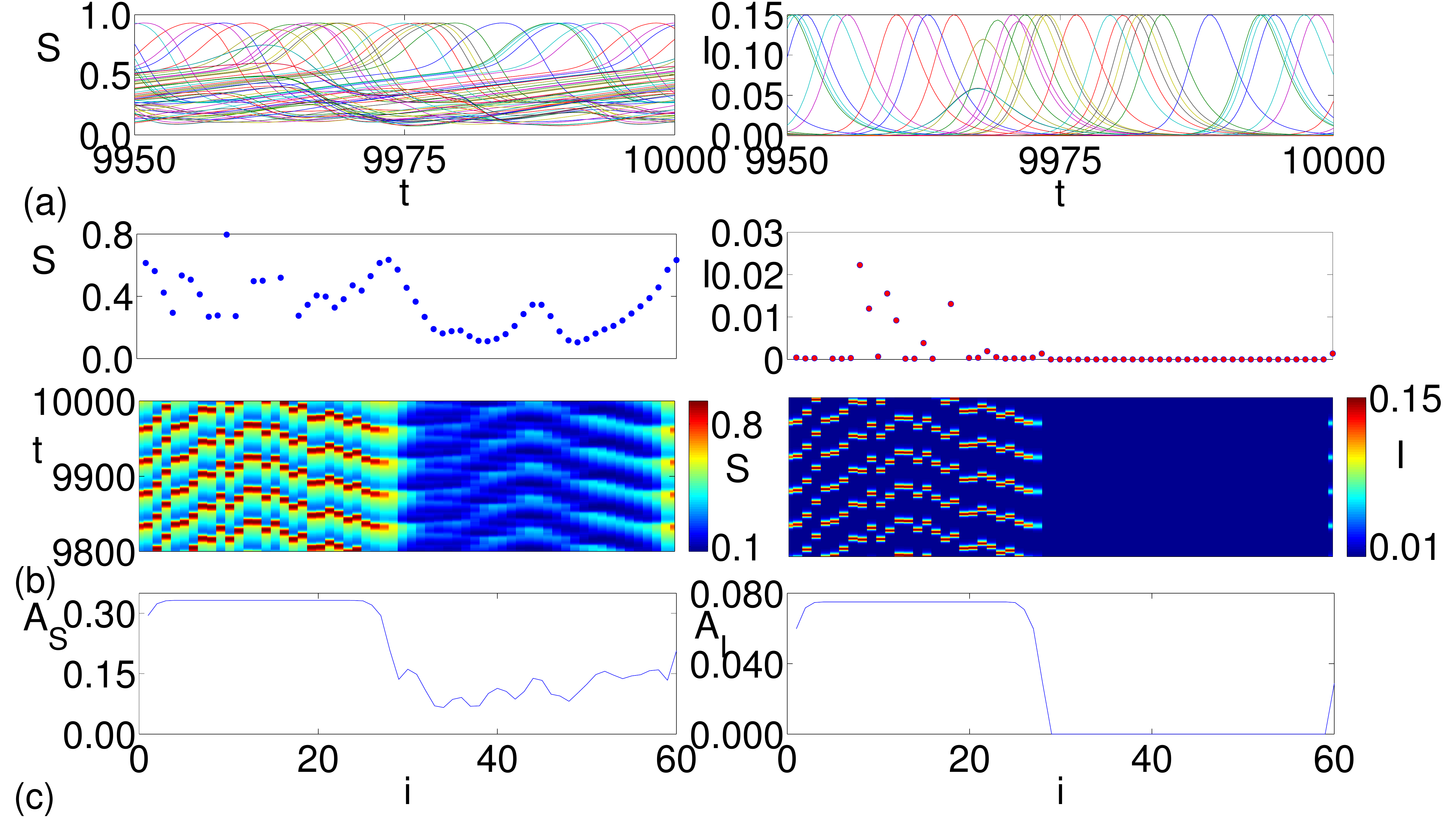}
	\caption{\label{fig:gauss_chimera} Amplitude chimera: time series, snapshots, space-time plots, and amplitude profiles for system~\eqref{eq:SI} with Gaussian coupling given by Eq.~\eqref{eq:Gauss_sigma}. Coupling parameters are $\tau = 34, d = 2$, and $\sigma = 0.7$, with other parameters as in Fig.~\ref{fig:re_lamb_vs_tau}.}
\end{figure}

\begin{figure}[ht!]
	\includegraphics[width=\linewidth]{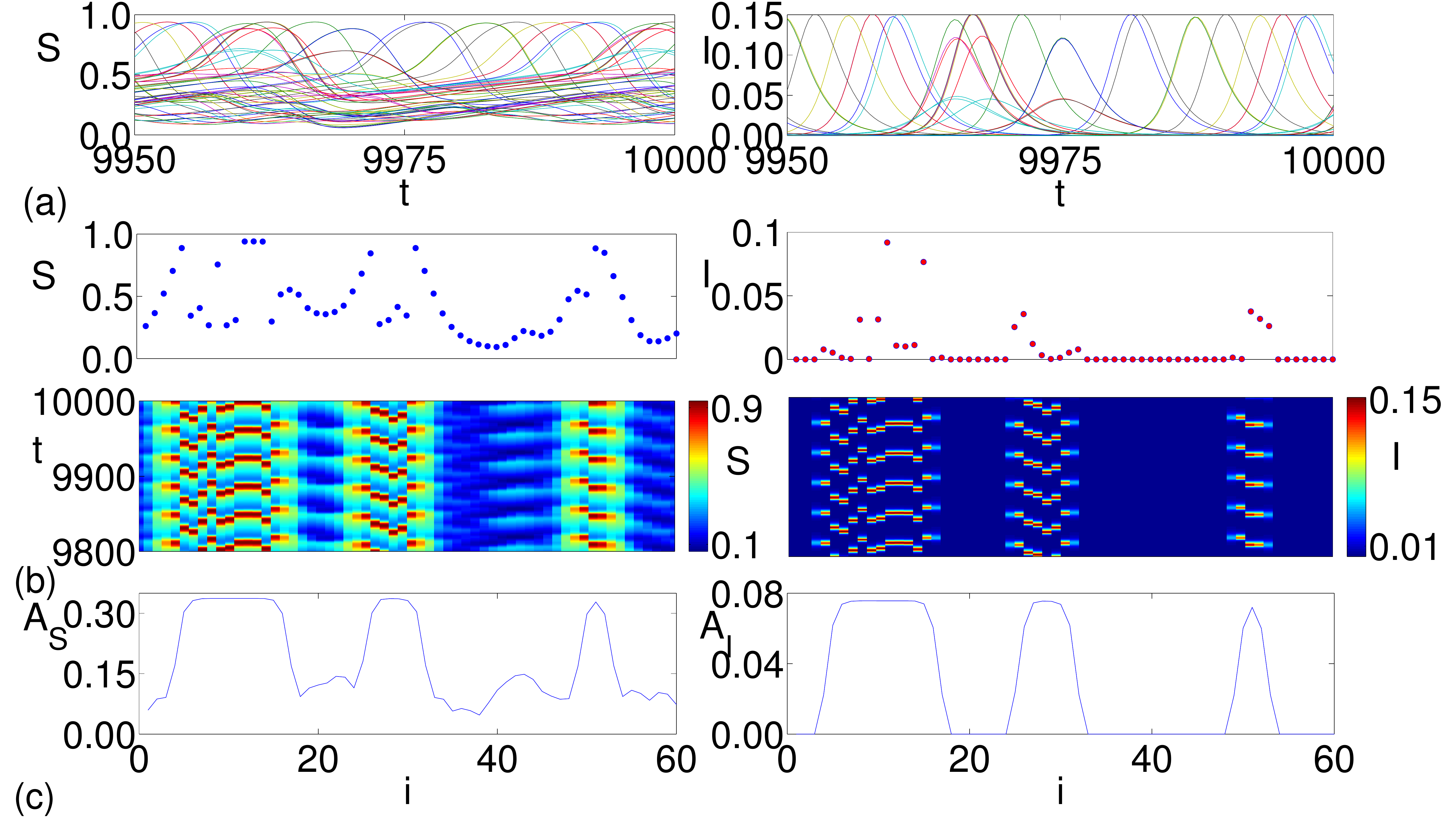}
	\caption{\label{fig:gauss_multi_chimera} Multi-headed amplitude chimera: time series, snapshots, space-time, and amplitude profile plots for system~\eqref{eq:SI} with Gaussian coupling given by Eq.~\eqref{eq:Gauss_sigma}. Coupling parameters are $\tau = 28, d = 2$, and $\sigma = 0.7$, with other parameters as in Fig.~\ref{fig:re_lamb_vs_tau}.}
\end{figure}

\begin{figure}[ht!]
	\includegraphics[width=\linewidth]{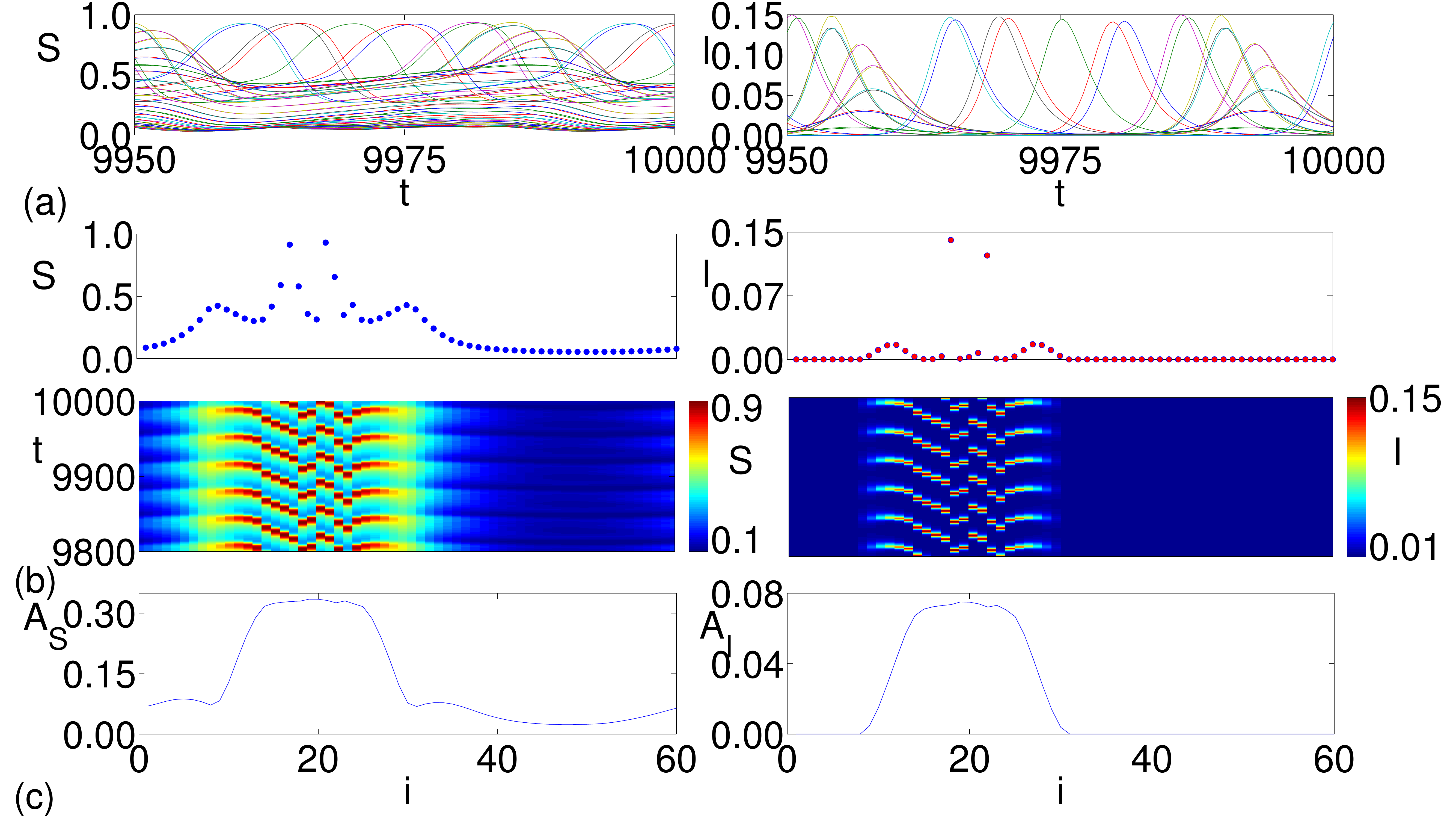}
	\caption{\label{fig:gauss_transition} Transition state between modulated profile and amplitude chimera: time series, snapshots, space-time plots, and amplitude profiles according to equations~\eqref{eq:SI} with Gaussian coupling given by Eq.~\eqref{eq:Gauss_sigma}. Coupling parameters are $\tau = 26, d = 8$, and $\sigma = 0.7$, with other parameters as in Fig.~\ref{fig:re_lamb_vs_tau}.}
\end{figure}

Below we present and discuss different patterns observed in the case of a non-local Gaussian coupling kernel~\eqref{eq:Gauss_sigma}. 
Figures~\ref{fig:gauss_modulated}-\ref{fig:gauss_transition} illustrate a modulated-amplitude profile, a solitary state, a traveling wave, (multi-headed) amplitude chimeras, and a transition state, respectively. To get a better insight into the dynamics, in each case the actual time series is plotted for all $N$ strains, accompanied by a snapshot at a fixed moment in time, a space-time plot, as well as plots of the average amplitude of oscillations for both dynamical variables. The amplitude is computed as the difference between maximum and minimum values of the respective variable for each strain.

Figure~\ref{fig:gauss_modulated} shows a regime where all strains oscillate with the same frequency and without phase shift, but with different amplitudes, as is clear from the space-time plots and the plots of the amplitude. Since many of the $I$ variables stay equal to zero in a manner similar to all-to-all coupling, while the frequency of oscillations is the same for $S$ variable for all oscillators, for $I$ variables it gets adjusted to the frequency of $S$ variables for those strains that do exhibit oscillations. The highest amplitude of oscillations occurs in the middle of modulated profile, suggesting the potential for amplitude rather than phase chimeras, but since the snapshot of the modulate profile is smooth, this state cannot be interpreted as a proper chimera state \cite{OME11,OME12}. From epidemiological perspective, this is an interesting state in that all non-zero strains follow synchronous oscillations, namely, they appear and disappear at the same time. On the other hand, infected fractions have substantially different magnitudes, which means that immunological interactions between strains results in some of them always being more dominant (i.e. having a significantly larger amplitude), whereas other strains are more suppressed, and this relation between different strains is repeated with every oscillation.

An exemplary case, where only a few strains exhibit oscillations of considerable amplitude, is shown in Fig.~\ref{fig:gauss_solitary} for a larger value of time delay $\tau=42$ and a smaller characteristic length $d=4$.

For small coupling strength $\sigma$ and narrow, that is, local, coupling kernels, we find a traveling wave pattern shown in Fig.~\ref{fig:gauss_wave}. This observation is important from a biological point of view, as it illustrates a regime of sequential strain dominance, which is often observed in epidemiological data \cite{REC09,BLY14}. 
%

In Fig.~\ref{fig:gauss_chimera} we illustrate the dynamical regime of an amplitude chimera. The different strains still oscillate with the same frequency, but in contrast to the previous pattern, the snapshots do not exhibit a smooth profile anymore but rather are represented by two different regions: coherent and incoherent. It is worth noting that whilst the coupling is still non-local, the amplitude chimera is observed even when the characteristic length of the coupling is quite small $(d=2)$. For the same parameter values but a smaller time delay, the system can also exhibit a multi-headed amplitude chimera, characterized by several coherent and incoherent regions with almost no variation in terms of frequency, but showing the amplitude profile typical for chimera states. An example of such state is shown in Fig.~\ref{fig:gauss_multi_chimera}.

A pattern of transition between an amplitude chimera and a modulated profile is demonstrated in Fig.~\ref{fig:gauss_transition}. Whilst there is an incoherent region in the middle of strain domain, the edges of the chimera have a smooth profile similar to that of the modulated profile, indicating that being a transition, this regime features the characteristics of both the chimera and the modulated profile. Similar to the amplitude chimera, the largest amplitude of oscillations for the transition state occurs in the incoherent regime. It should be noted that transition states can be found for a whole range of parameter values between modulated profile and amplitude chimeras, making them closer in terms of dynamics to either of those states.

\begin{figure}[ht!]
	\includegraphics[width=\linewidth]{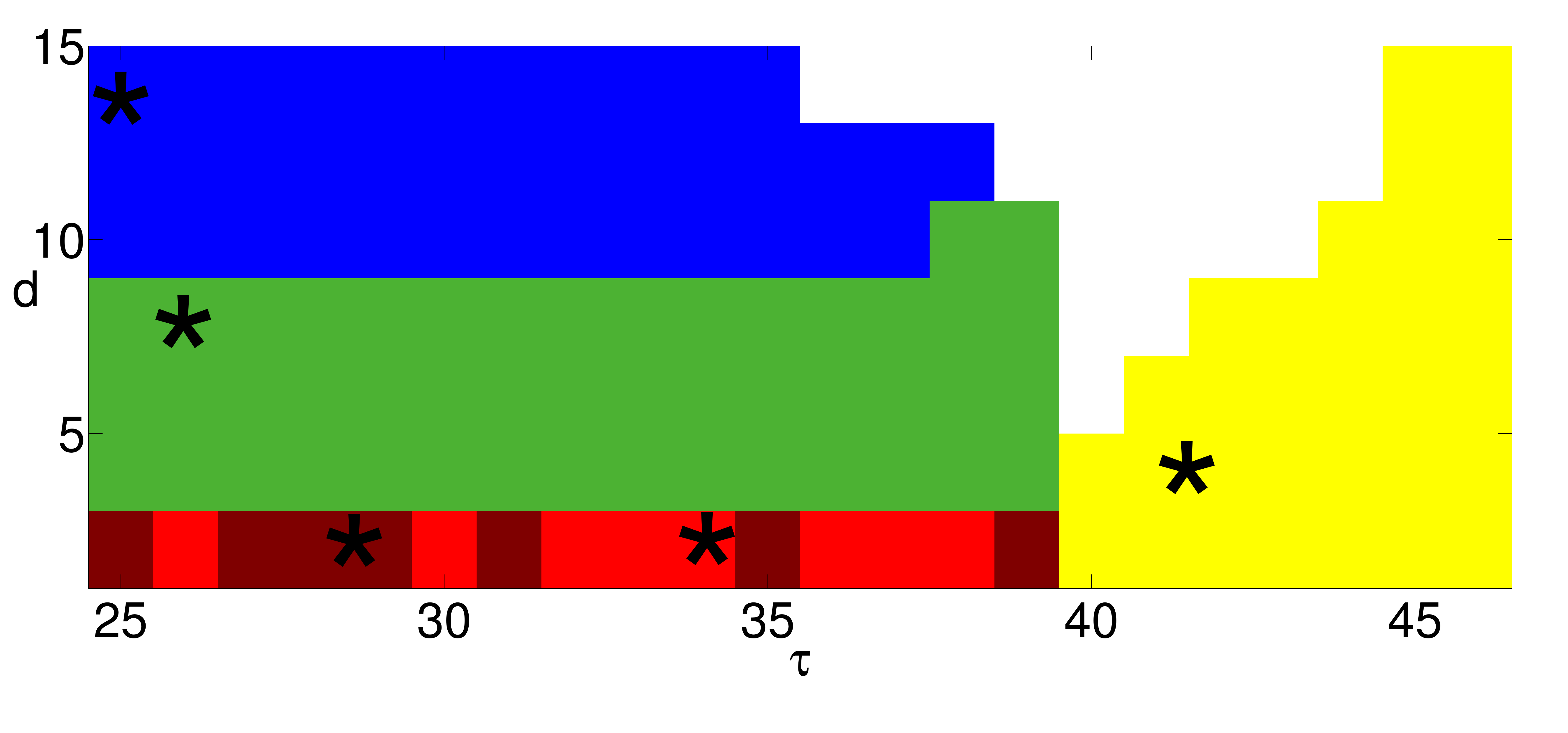}
	\caption{\label{fig:gauss_scan} Parameter regions of different dynamical regimes for the Gaussian coupling~\eqref{eq:Gauss_sigma} depending on the time delay $\tau$ and the standard deviation $d$. The blue, green, (dark) red, white and yellow regions refer to states of modulated profile, transition to chimera, (multi-headed) chimera, transition between modulated and solitary states, and solitary states, respectively. The markers $*$ indicate the parameter combinations used in Figs.~\ref{fig:gauss_modulated}, \ref{fig:gauss_solitary}, \ref{fig:gauss_chimera}, \ref{fig:gauss_multi_chimera}, and \ref{fig:gauss_transition}. Other parameters as in Fig.~\ref{fig:gauss_modulated}.}
\end{figure}

Figure~\ref{fig:gauss_scan} provides a summary of different dynamical states that can be observed in the system~\eqref{eq:SI} depending on the time delay $\tau$ and the cross-immunity length scale $d$. Larger values of the cross-immunity length scale, i.e. broader coupling kernels, are associated with modulated amplitude profiles, while, surprisingly, chimera states (single- and multi-headed) are found for narrower, i.e. more local, coupling kernels. Solitary states in which infections with only a single strain are present, can occur for any lengths of cross-immunity $d$, provided the time delay $\tau$ is sufficiently large.

\begin{figure}[ht!]
	\includegraphics[width=\linewidth]{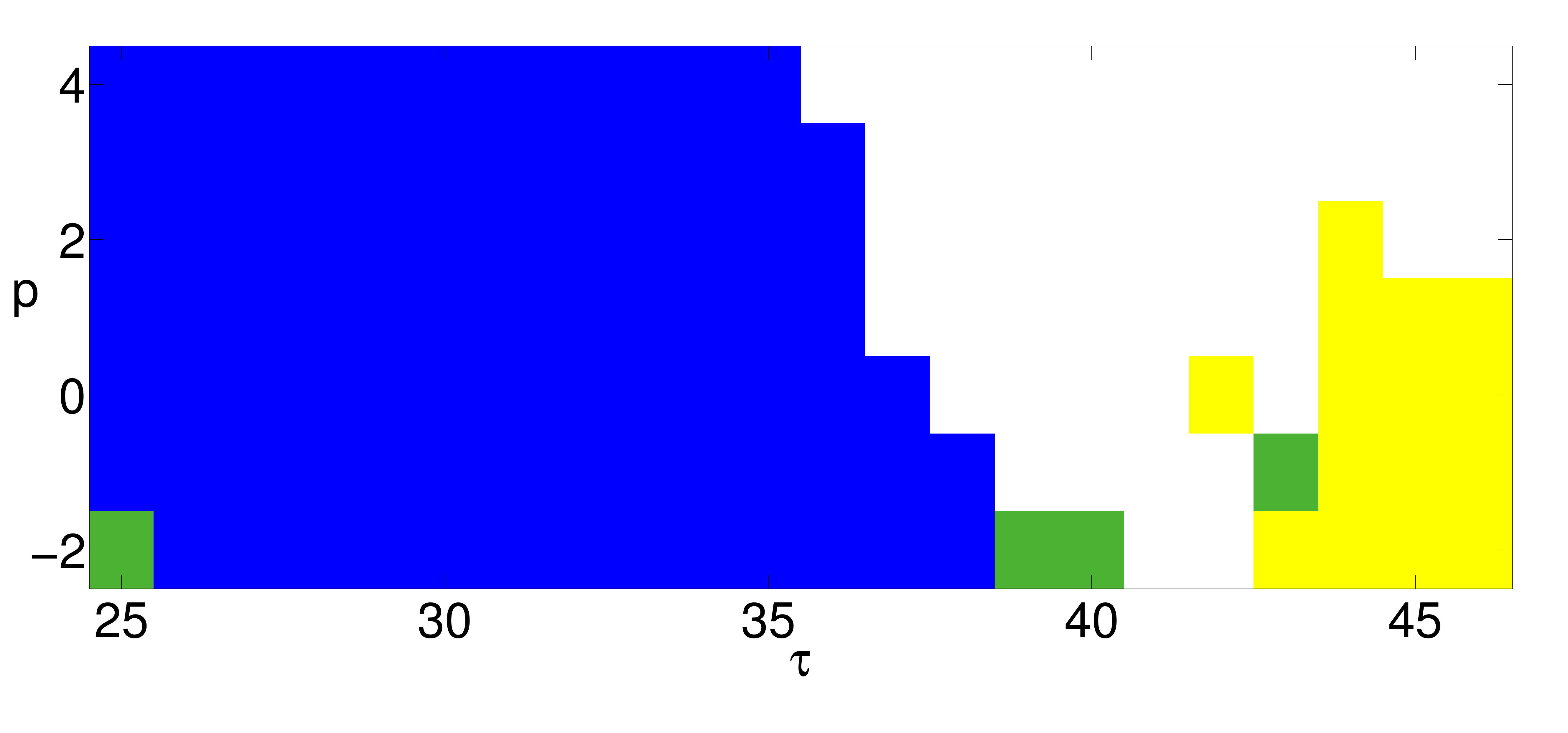}
	\caption{\label{fig:cosine_scan} Parameter regions of different dynamical regimes for the cosine coupling~\eqref{eq:cos_coup},\eqref{eq:cos_coup_dp} depending on the time delay $\tau$ and coupling parameter $p$. The blue, green, white and yellow regions refer to states of modulated profile, transition to chimera, transition between modulated and solitary states, and solitary states, respectively. Other parameters as in Fig.~\ref{fig:gauss_modulated}.}
\end{figure}

We have also performed extensive simulations for the case of cosine kernel~\eqref{eq:cos_coup}, and a summary of results is shown in Fig.~\ref{fig:cosine_scan}. Unlike the Gaussian kernel, in this case only modulated profiles, solitary states, and transition states are observed, while traveling waves and amplitude chimeras were never found. The most likely explanation for this lies in the fact that amplitude chimeras are associated with quite narrow Gaussian kernel (as described by small values of $d$), whereas for the biologically feasible values of parameter $p\in[-2,4]$, the distribution~\eqref{eq:cos_coup} is quite broad. In fact, Figure~\ref{fig:cos_func} suggests that the narrowest width of the cosine distribution corresponds to $p=-2$, which, for a system of $N=60$ strains is equivalent to $d=12$, and for large values of $p$, the coupling is very broad, making it more similar to the situation described by an all-to-all coupling. As a result, the dynamics is dominated by modulated amplitude profiles for smaller durations of temporary immunity, and by solitary states with single-strain dynamics for larger values of the time delay.

\section{Discussion}
\label{sec:conclusion}

In this paper we have studied an important question about the range of dynamical behaviors that can be exhibited by multi-strain epidemic models with temporary immunity and various types of cross-immunity. Whilst the time delay associated with temporary immunity provides a simple mechanism supporting stable oscillations in the susceptible and infected populations for individual disease strains, this dynamics undergoes major changes under the influence of long-range coupling. Under assumption of all-to-all coupling, the system settles on the dynamical regime of solitary states, or single-strain oscillations, where infections with only one strain are present, while all other strains remain equal to zero. Interestingly, the system approaches such a state for random initial conditions, suggesting that this state is, in fact, a stable invariant manifold of the model, which dynamically represents a symmetry-breaking suppression of oscillations. The complete symmetry between strains means that the surviving strain is determined purely by the initial conditions, and for the same parameter values, all other strains are equally possible.

For the case of Gaussian cross-immunity kernel, the model exhibits a wide range of dynamical scenarios that include solitary states, traveling waves, and, most interestingly, single- and multi-headed amplitude chimeras, characterized by some groups of strains oscillating coherently, while others are performing incoherent oscillations. Whilst the cosine kernel is also non-local, by virtue of being very broad, the range of different behaviors for this kernel is smaller and is more reminiscent of the case of all-to-all coupling. The fact that chimera states were observed only for sufficiently narrow cross-immunity kernels suggests that in epidemiological data these types of solutions would only be observed in the cases where individual strains or serotypes elicit cross-reactive immune responses against very genetically similar strains. For multi-strain diseases with a wide antigenic repertoire, chimera states could be interpreted as dynamical regimes where a number of closely immunologically related strains appear to have similar dynamics and show up concurrently, while other strains have irregular and unsynchronized oscillations. Understanding parameter regimes that result in chimera states can provide useful insights for design and deployment of multi-valent vaccines.


\begin{acknowledgments}
PH and JB acknowledge support by Deutsche Forschungsgemeinschaft in the framework of the
Collaborative Research Center SFB 910. 
YK and KB acknowledge the hospitality of the Institute of Theoretical Physics, TU Berlin, where part of this work was completed.
\end{acknowledgments}


\begin{thebibliography}{62}%
\makeatletter
\providecommand \@ifxundefined [1]{%
 \@ifx{#1\undefined}
}%
\providecommand \@ifnum [1]{%
 \ifnum #1\expandafter \@firstoftwo
 \else \expandafter \@secondoftwo
 \fi
}%
\providecommand \@ifx [1]{%
 \ifx #1\expandafter \@firstoftwo
 \else \expandafter \@secondoftwo
 \fi
}%
\providecommand \natexlab [1]{#1}%
\providecommand \enquote  [1]{``#1''}%
\providecommand \bibnamefont  [1]{#1}%
\providecommand \bibfnamefont [1]{#1}%
\providecommand \citenamefont [1]{#1}%
\providecommand \href@noop [0]{\@secondoftwo}%
\providecommand \href [0]{\begingroup \@sanitize@url \@href}%
\providecommand \@href[1]{\@@startlink{#1}\@@href}%
\providecommand \@@href[1]{\endgroup#1\@@endlink}%
\providecommand \@sanitize@url [0]{\catcode `\\12\catcode `\$12\catcode
  `\&12\catcode `\#12\catcode `\^12\catcode `\_12\catcode `\%12\relax}%
\providecommand \@@startlink[1]{}%
\providecommand \@@endlink[0]{}%
\providecommand \url  [0]{\begingroup\@sanitize@url \@url }%
\providecommand \@url [1]{\endgroup\@href {#1}{\urlprefix }}%
\providecommand \urlprefix  [0]{URL }%
\providecommand \Eprint [0]{\href }%
\providecommand \doibase [0]{http://dx.doi.org/}%
\providecommand \selectlanguage [0]{\@gobble}%
\providecommand \bibinfo  [0]{\@secondoftwo}%
\providecommand \bibfield  [0]{\@secondoftwo}%
\providecommand \translation [1]{[#1]}%
\providecommand \BibitemOpen [0]{}%
\providecommand \bibitemStop [0]{}%
\providecommand \bibitemNoStop [0]{.\EOS\space}%
\providecommand \EOS [0]{\spacefactor3000\relax}%
\providecommand \BibitemShut  [1]{\csname bibitem#1\endcsname}%
\let\auto@bib@innerbib\@empty
\bibitem [{\citenamefont {Kuramoto}\ and\ \citenamefont
  {Battogtokh}(2002)}]{KUR02a}%
  \BibitemOpen
  \bibfield  {author} {\bibinfo {author} {\bibfnamefont {Y.}~\bibnamefont
  {Kuramoto}}\ and\ \bibinfo {author} {\bibfnamefont {D.}~\bibnamefont
  {Battogtokh}},\ }\href@noop {} {\bibfield  {journal} {\bibinfo  {journal}
  {Nonlin. Phen. in Complex Sys.}\ }\textbf {\bibinfo {volume} {5}},\ \bibinfo
  {pages} {380} (\bibinfo {year} {2002})}\BibitemShut {NoStop}%
\bibitem [{\citenamefont {Abrams}\ and\ \citenamefont
  {Strogatz}(2004)}]{ABR04}%
  \BibitemOpen
  \bibfield  {author} {\bibinfo {author} {\bibfnamefont {D.~M.}\ \bibnamefont
  {Abrams}}\ and\ \bibinfo {author} {\bibfnamefont {S.~H.}\ \bibnamefont
  {Strogatz}},\ }\href {\doibase 10.1103/physrevlett.93.174102} {\bibfield
  {journal} {\bibinfo  {journal} {Phys.~Rev.~Lett.}\ }\textbf {\bibinfo
  {volume} {93}},\ \bibinfo {pages} {174102} (\bibinfo {year}
  {2004})}\BibitemShut {NoStop}%
\bibitem [{\citenamefont {Lazarides}, \citenamefont {Neofotistos},\ and\
  \citenamefont {Tsironis}(2015)}]{LAZ15}%
  \BibitemOpen
  \bibfield  {author} {\bibinfo {author} {\bibfnamefont {N.}~\bibnamefont
  {Lazarides}}, \bibinfo {author} {\bibfnamefont {G.}~\bibnamefont
  {Neofotistos}}, \ and\ \bibinfo {author} {\bibfnamefont {G.~P.}\ \bibnamefont
  {Tsironis}},\ }\href@noop {} {\bibfield  {journal} {\bibinfo  {journal}
  {Phys.~Rev.~B}\ }\textbf {\bibinfo {volume} {91}},\ \bibinfo {pages} {054303}
  (\bibinfo {year} {2015})}\BibitemShut {NoStop}%
\bibitem [{\citenamefont {Bastidas}\ \emph {et~al.}(2015)\citenamefont
  {Bastidas}, \citenamefont {Omelchenko}, \citenamefont {Zakharova},
  \citenamefont {Sch{\"o}ll},\ and\ \citenamefont {Brandes}}]{BAS15}%
  \BibitemOpen
  \bibfield  {author} {\bibinfo {author} {\bibfnamefont {V.}~\bibnamefont
  {Bastidas}}, \bibinfo {author} {\bibfnamefont {I.}~\bibnamefont
  {Omelchenko}}, \bibinfo {author} {\bibfnamefont {A.}~\bibnamefont
  {Zakharova}}, \bibinfo {author} {\bibfnamefont {E.}~\bibnamefont
  {Sch{\"o}ll}}, \ and\ \bibinfo {author} {\bibfnamefont {T.}~\bibnamefont
  {Brandes}},\ }\href {\doibase 10.1103/physreve.92.062924} {\bibfield
  {journal} {\bibinfo  {journal} {Phys. Rev. E}\ }\textbf {\bibinfo {volume}
  {92}},\ \bibinfo {pages} {062924} (\bibinfo {year} {2015})}\BibitemShut
  {NoStop}%
\bibitem [{\citenamefont {Gambuzza}\ \emph {et~al.}(2014)\citenamefont
  {Gambuzza}, \citenamefont {Buscarino}, \citenamefont {Chessari},
  \citenamefont {Fortuna}, \citenamefont {Meucci},\ and\ \citenamefont
  {Frasca}}]{GAM14}%
  \BibitemOpen
  \bibfield  {author} {\bibinfo {author} {\bibfnamefont {L.~V.}\ \bibnamefont
  {Gambuzza}}, \bibinfo {author} {\bibfnamefont {A.}~\bibnamefont {Buscarino}},
  \bibinfo {author} {\bibfnamefont {S.}~\bibnamefont {Chessari}}, \bibinfo
  {author} {\bibfnamefont {L.}~\bibnamefont {Fortuna}}, \bibinfo {author}
  {\bibfnamefont {R.}~\bibnamefont {Meucci}}, \ and\ \bibinfo {author}
  {\bibfnamefont {M.}~\bibnamefont {Frasca}},\ }\href {\doibase
  10.1103/physreve.90.032905} {\bibfield  {journal} {\bibinfo  {journal} {Phys.
  Rev. E}\ }\textbf {\bibinfo {volume} {90}},\ \bibinfo {pages} {032905}
  (\bibinfo {year} {2014})}\BibitemShut {NoStop}%
\bibitem [{\citenamefont {Rattenborg}, \citenamefont {Amlaner},\ and\
  \citenamefont {Lima}(2000)}]{RAT00}%
  \BibitemOpen
  \bibfield  {author} {\bibinfo {author} {\bibfnamefont {N.~C.}\ \bibnamefont
  {Rattenborg}}, \bibinfo {author} {\bibfnamefont {C.~J.}\ \bibnamefont
  {Amlaner}}, \ and\ \bibinfo {author} {\bibfnamefont {S.~L.}\ \bibnamefont
  {Lima}},\ }\href@noop {} {\bibfield  {journal} {\bibinfo  {journal}
  {Neurosci. Biobehav. Rev.}\ }\textbf {\bibinfo {volume} {24}},\ \bibinfo
  {pages} {817} (\bibinfo {year} {2000})}\BibitemShut {NoStop}%
\bibitem [{\citenamefont {Filatrella}, \citenamefont {Nielsen},\ and\
  \citenamefont {Pedersen}(2008)}]{FIL08a}%
  \BibitemOpen
  \bibfield  {author} {\bibinfo {author} {\bibfnamefont {G.}~\bibnamefont
  {Filatrella}}, \bibinfo {author} {\bibfnamefont {A.~H.}\ \bibnamefont
  {Nielsen}}, \ and\ \bibinfo {author} {\bibfnamefont {N.~F.}\ \bibnamefont
  {Pedersen}},\ }\href@noop {} {\bibfield  {journal} {\bibinfo  {journal} {Eur.
  Phys.~J.~B}\ }\textbf {\bibinfo {volume} {61}},\ \bibinfo {pages} {485}
  (\bibinfo {year} {2008})}\BibitemShut {NoStop}%
\bibitem [{\citenamefont {Martens}\ \emph {et~al.}(2013)\citenamefont
  {Martens}, \citenamefont {Thutupalli}, \citenamefont {Fourriere},\ and\
  \citenamefont {Hallatschek}}]{MAR13}%
  \BibitemOpen
  \bibfield  {author} {\bibinfo {author} {\bibfnamefont {E.~A.}\ \bibnamefont
  {Martens}}, \bibinfo {author} {\bibfnamefont {S.}~\bibnamefont {Thutupalli}},
  \bibinfo {author} {\bibfnamefont {A.}~\bibnamefont {Fourriere}}, \ and\
  \bibinfo {author} {\bibfnamefont {O.}~\bibnamefont {Hallatschek}},\ }\href
  {\doibase 10.1073/pnas.1302880110} {\bibfield  {journal} {\bibinfo  {journal}
  {Proc. Natl. Acad. Sci. USA}\ }\textbf {\bibinfo {volume} {110}},\ \bibinfo
  {pages} {10563} (\bibinfo {year} {2013})}\BibitemShut {NoStop}%
\bibitem [{\citenamefont {Motter}\ \emph {et~al.}(2013)\citenamefont {Motter},
  \citenamefont {Myers}, \citenamefont {Anghel},\ and\ \citenamefont
  {Nishikawa}}]{MOT13a}%
  \BibitemOpen
  \bibfield  {author} {\bibinfo {author} {\bibfnamefont {A.~E.}\ \bibnamefont
  {Motter}}, \bibinfo {author} {\bibfnamefont {S.~A.}\ \bibnamefont {Myers}},
  \bibinfo {author} {\bibfnamefont {M.}~\bibnamefont {Anghel}}, \ and\ \bibinfo
  {author} {\bibfnamefont {T.}~\bibnamefont {Nishikawa}},\ }\href {\doibase
  doi:10.1038/nphys2535} {\bibfield  {journal} {\bibinfo  {journal} {Nature
  Phys.}\ }\textbf {\bibinfo {volume} {9}},\ \bibinfo {pages} {191} (\bibinfo
  {year} {2013})}\BibitemShut {NoStop}%
\bibitem [{\citenamefont {Omelchenko}\ \emph {et~al.}(2015)\citenamefont
  {Omelchenko}, \citenamefont {Provata}, \citenamefont {Hizanidis},
  \citenamefont {Sch{\"o}ll},\ and\ \citenamefont {H{\"o}vel}}]{OME15}%
  \BibitemOpen
  \bibfield  {author} {\bibinfo {author} {\bibfnamefont {I.}~\bibnamefont
  {Omelchenko}}, \bibinfo {author} {\bibfnamefont {A.}~\bibnamefont {Provata}},
  \bibinfo {author} {\bibfnamefont {J.}~\bibnamefont {Hizanidis}}, \bibinfo
  {author} {\bibfnamefont {E.}~\bibnamefont {Sch{\"o}ll}}, \ and\ \bibinfo
  {author} {\bibfnamefont {P.}~\bibnamefont {H{\"o}vel}},\ }\href {\doibase
  10.1103/physreve.91.022917} {\bibfield  {journal} {\bibinfo  {journal} {Phys.
  Rev. E}\ }\textbf {\bibinfo {volume} {91}},\ \bibinfo {pages} {022917}
  (\bibinfo {year} {2015})}\BibitemShut {NoStop}%
\bibitem [{\citenamefont {Martens}, \citenamefont {Panaggio},\ and\
  \citenamefont {Abrams}(2016)}]{MAR16c}%
  \BibitemOpen
  \bibfield  {author} {\bibinfo {author} {\bibfnamefont {E.~A.}\ \bibnamefont
  {Martens}}, \bibinfo {author} {\bibfnamefont {M.~J.}\ \bibnamefont
  {Panaggio}}, \ and\ \bibinfo {author} {\bibfnamefont {D.~M.}\ \bibnamefont
  {Abrams}},\ }\href@noop {} {\bibfield  {journal} {\bibinfo  {journal} {New~J.
  Phys.}\ }\textbf {\bibinfo {volume} {18}},\ \bibinfo {pages} {022002}
  (\bibinfo {year} {2016})}\BibitemShut {NoStop}%
\bibitem [{\citenamefont {Panaggio}\ \emph {et~al.}(2016)\citenamefont
  {Panaggio}, \citenamefont {Abrams}, \citenamefont {Ashwin},\ and\
  \citenamefont {Laing}}]{PAN16c}%
  \BibitemOpen
  \bibfield  {author} {\bibinfo {author} {\bibfnamefont {M.~J.}\ \bibnamefont
  {Panaggio}}, \bibinfo {author} {\bibfnamefont {D.~M.}\ \bibnamefont
  {Abrams}}, \bibinfo {author} {\bibfnamefont {P.}~\bibnamefont {Ashwin}}, \
  and\ \bibinfo {author} {\bibfnamefont {C.~R.}\ \bibnamefont {Laing}},\ }\href
  {\doibase 10.1103/physreve.93.012218} {\bibfield  {journal} {\bibinfo
  {journal} {Phys. Rev. E}\ }\textbf {\bibinfo {volume} {93}},\ \bibinfo
  {pages} {012218} (\bibinfo {year} {2016})}\BibitemShut {NoStop}%
\bibitem [{\citenamefont {Wolfrum}\ and\ \citenamefont
  {Omel'chenko}(2011)}]{WOL11}%
  \BibitemOpen
  \bibfield  {author} {\bibinfo {author} {\bibfnamefont {M.}~\bibnamefont
  {Wolfrum}}\ and\ \bibinfo {author} {\bibfnamefont {O.~E.}\ \bibnamefont
  {Omel'chenko}},\ }\href {\doibase 10.1103/physreve.84.015201} {\bibfield
  {journal} {\bibinfo  {journal} {Phys. Rev. E}\ }\textbf {\bibinfo {volume}
  {84}},\ \bibinfo {pages} {015201} (\bibinfo {year} {2011})}\BibitemShut
  {NoStop}%
\bibitem [{\citenamefont {Olmi}(2015)}]{OLM15a}%
  \BibitemOpen
  \bibfield  {author} {\bibinfo {author} {\bibfnamefont {S.}~\bibnamefont
  {Olmi}},\ }\href {\doibase 10.1063/1.4938734} {\bibfield  {journal} {\bibinfo
   {journal} {Chaos}\ }\textbf {\bibinfo {volume} {25}},\ \bibinfo {pages}
  {123125} (\bibinfo {year} {2015})}\BibitemShut {NoStop}%
\bibitem [{\citenamefont {Olmi}\ \emph {et~al.}(2015)\citenamefont {Olmi},
  \citenamefont {Martens}, \citenamefont {Thutupalli},\ and\ \citenamefont
  {Torcini}}]{OLM15}%
  \BibitemOpen
  \bibfield  {author} {\bibinfo {author} {\bibfnamefont {S.}~\bibnamefont
  {Olmi}}, \bibinfo {author} {\bibfnamefont {E.~A.}\ \bibnamefont {Martens}},
  \bibinfo {author} {\bibfnamefont {S.}~\bibnamefont {Thutupalli}}, \ and\
  \bibinfo {author} {\bibfnamefont {A.}~\bibnamefont {Torcini}},\ }\href
  {\doibase 10.1103/physreve.92.030901} {\bibfield  {journal} {\bibinfo
  {journal} {Phys. Rev. E}\ }\textbf {\bibinfo {volume} {92}},\ \bibinfo
  {pages} {030901(R)} (\bibinfo {year} {2015})}\BibitemShut {NoStop}%
\bibitem [{\citenamefont {Nkomo}, \citenamefont {Tinsley},\ and\ \citenamefont
  {Showalter}(2013)}]{NKO13}%
  \BibitemOpen
  \bibfield  {author} {\bibinfo {author} {\bibfnamefont {S.}~\bibnamefont
  {Nkomo}}, \bibinfo {author} {\bibfnamefont {M.~R.}\ \bibnamefont {Tinsley}},
  \ and\ \bibinfo {author} {\bibfnamefont {K.}~\bibnamefont {Showalter}},\
  }\href {\doibase 10.1103/physrevlett.110.244102} {\bibfield  {journal}
  {\bibinfo  {journal} {Phys. Rev. Lett.}\ }\textbf {\bibinfo {volume} {110}},\
  \bibinfo {pages} {244102} (\bibinfo {year} {2013})}\BibitemShut {NoStop}%
\bibitem [{\citenamefont {Hagerstrom}\ \emph {et~al.}(2012)\citenamefont
  {Hagerstrom}, \citenamefont {Murphy}, \citenamefont {Roy}, \citenamefont
  {H{\"o}vel}, \citenamefont {Omelchenko},\ and\ \citenamefont
  {Sch{\"o}ll}}]{HAG12}%
  \BibitemOpen
  \bibfield  {author} {\bibinfo {author} {\bibfnamefont {A.~M.}\ \bibnamefont
  {Hagerstrom}}, \bibinfo {author} {\bibfnamefont {T.~E.}\ \bibnamefont
  {Murphy}}, \bibinfo {author} {\bibfnamefont {R.}~\bibnamefont {Roy}},
  \bibinfo {author} {\bibfnamefont {P.}~\bibnamefont {H{\"o}vel}}, \bibinfo
  {author} {\bibfnamefont {I.}~\bibnamefont {Omelchenko}}, \ and\ \bibinfo
  {author} {\bibfnamefont {E.}~\bibnamefont {Sch{\"o}ll}},\ }\href {\doibase
  10.1038/nphys2372} {\bibfield  {journal} {\bibinfo  {journal} {Nature Phys.}\
  }\textbf {\bibinfo {volume} {8}},\ \bibinfo {pages} {658} (\bibinfo {year}
  {2012})}\BibitemShut {NoStop}%
\bibitem [{\citenamefont {Larger}, \citenamefont {Penkovsky},\ and\
  \citenamefont {Maistrenko}(2015)}]{LAR15}%
  \BibitemOpen
  \bibfield  {author} {\bibinfo {author} {\bibfnamefont {L.}~\bibnamefont
  {Larger}}, \bibinfo {author} {\bibfnamefont {B.}~\bibnamefont {Penkovsky}}, \
  and\ \bibinfo {author} {\bibfnamefont {Y.}~\bibnamefont {Maistrenko}},\
  }\href@noop {} {\bibfield  {journal} {\bibinfo  {journal} {Nature Commun.}\
  }\textbf {\bibinfo {volume} {6}},\ \bibinfo {pages} {7752} (\bibinfo {year}
  {2015})}\BibitemShut {NoStop}%
\bibitem [{\citenamefont {Larger}, \citenamefont {Penkovsky},\ and\
  \citenamefont {Maistrenko}(2013)}]{LAR13}%
  \BibitemOpen
  \bibfield  {author} {\bibinfo {author} {\bibfnamefont {L.}~\bibnamefont
  {Larger}}, \bibinfo {author} {\bibfnamefont {B.}~\bibnamefont {Penkovsky}}, \
  and\ \bibinfo {author} {\bibfnamefont {Y.}~\bibnamefont {Maistrenko}},\
  }\href {\doibase 10.1103/physrevlett.111.054103} {\bibfield  {journal}
  {\bibinfo  {journal} {Phys. Rev. Lett.}\ }\textbf {\bibinfo {volume} {111}},\
  \bibinfo {pages} {054103} (\bibinfo {year} {2013})}\BibitemShut {NoStop}%
\bibitem [{\citenamefont {Panaggio}\ and\ \citenamefont
  {Abrams}(2015)}]{PAN15}%
  \BibitemOpen
  \bibfield  {author} {\bibinfo {author} {\bibfnamefont {M.~J.}\ \bibnamefont
  {Panaggio}}\ and\ \bibinfo {author} {\bibfnamefont {D.~M.}\ \bibnamefont
  {Abrams}},\ }\href {\doibase 10.1088/0951-7715/28/3/r67} {\bibfield
  {journal} {\bibinfo  {journal} {Nonlinearity}\ }\textbf {\bibinfo {volume}
  {28}},\ \bibinfo {pages} {R67} (\bibinfo {year} {2015})}\BibitemShut
  {NoStop}%
\bibitem [{\citenamefont {Omelchenko}\ \emph {et~al.}(2013)\citenamefont
  {Omelchenko}, \citenamefont {Omel'chenko}, \citenamefont {H{\"o}vel},\ and\
  \citenamefont {Sch{\"o}ll}}]{OME13}%
  \BibitemOpen
  \bibfield  {author} {\bibinfo {author} {\bibfnamefont {I.}~\bibnamefont
  {Omelchenko}}, \bibinfo {author} {\bibfnamefont {O.~E.}\ \bibnamefont
  {Omel'chenko}}, \bibinfo {author} {\bibfnamefont {P.}~\bibnamefont
  {H{\"o}vel}}, \ and\ \bibinfo {author} {\bibfnamefont {E.}~\bibnamefont
  {Sch{\"o}ll}},\ }\href {\doibase 10.1103/physrevlett.110.224101} {\bibfield
  {journal} {\bibinfo  {journal} {Phys. Rev. Lett.}\ }\textbf {\bibinfo
  {volume} {110}},\ \bibinfo {pages} {224101} (\bibinfo {year}
  {2013})}\BibitemShut {NoStop}%
\bibitem [{\citenamefont {Wang}\ and\ \citenamefont {Li}(2011)}]{WAN11f}%
  \BibitemOpen
  \bibfield  {author} {\bibinfo {author} {\bibfnamefont {H.}~\bibnamefont
  {Wang}}\ and\ \bibinfo {author} {\bibfnamefont {X.}~\bibnamefont {Li}},\
  }\href {\doibase 10.1103/physreve.83.066214} {\bibfield  {journal} {\bibinfo
  {journal} {Phys. Rev. E}\ }\textbf {\bibinfo {volume} {83}},\ \bibinfo
  {pages} {066214} (\bibinfo {year} {2011})}\BibitemShut {NoStop}%
\bibitem [{\citenamefont {Sethia}, \citenamefont {Sen},\ and\ \citenamefont
  {Johnston}(2013)}]{SET13}%
  \BibitemOpen
  \bibfield  {author} {\bibinfo {author} {\bibfnamefont {G.~C.}\ \bibnamefont
  {Sethia}}, \bibinfo {author} {\bibfnamefont {A.}~\bibnamefont {Sen}}, \ and\
  \bibinfo {author} {\bibfnamefont {G.~L.}\ \bibnamefont {Johnston}},\
  }\href@noop {} {\bibfield  {journal} {\bibinfo  {journal} {Phys. Rev. E}\
  }\textbf {\bibinfo {volume} {88}},\ \bibinfo {pages} {042917} (\bibinfo
  {year} {2013})}\BibitemShut {NoStop}%
\bibitem [{\citenamefont {Ulonska}\ \emph {et~al.}(2016)\citenamefont
  {Ulonska}, \citenamefont {Omelchenko}, \citenamefont {Zakharova},\ and\
  \citenamefont {Sch{\"o}ll}}]{ULO16}%
  \BibitemOpen
  \bibfield  {author} {\bibinfo {author} {\bibfnamefont {S.}~\bibnamefont
  {Ulonska}}, \bibinfo {author} {\bibfnamefont {I.}~\bibnamefont {Omelchenko}},
  \bibinfo {author} {\bibfnamefont {A.}~\bibnamefont {Zakharova}}, \ and\
  \bibinfo {author} {\bibfnamefont {E.}~\bibnamefont {Sch{\"o}ll}},\
  }\href@noop {} {\bibfield  {journal} {\bibinfo  {journal} {Chaos}\ }\textbf
  {\bibinfo {volume} {26}},\ \bibinfo {pages} {094825} (\bibinfo {year}
  {2016})}\BibitemShut {NoStop}%
\bibitem [{\citenamefont {Tsigkri-DeSmedt}\ \emph {et~al.}(2016)\citenamefont
  {Tsigkri-DeSmedt}, \citenamefont {Hizanidis}, \citenamefont {H{\"o}vel},\
  and\ \citenamefont {Provata}}]{TSI16}%
  \BibitemOpen
  \bibfield  {author} {\bibinfo {author} {\bibfnamefont {N.~D.}\ \bibnamefont
  {Tsigkri-DeSmedt}}, \bibinfo {author} {\bibfnamefont {J.}~\bibnamefont
  {Hizanidis}}, \bibinfo {author} {\bibfnamefont {P.}~\bibnamefont
  {H{\"o}vel}}, \ and\ \bibinfo {author} {\bibfnamefont {A.}~\bibnamefont
  {Provata}},\ }\href {\doibase 10.1140/epjst/e2016-02661-4} {\bibfield
  {journal} {\bibinfo  {journal} {Eur. Phys.~J.~ST}\ }\textbf {\bibinfo
  {volume} {225}},\ \bibinfo {pages} {1149} (\bibinfo {year}
  {2016})}\BibitemShut {NoStop}%
\bibitem [{\citenamefont {Zakharova}, \citenamefont {Kapeller},\ and\
  \citenamefont {Sch{\"o}ll}(2016)}]{ZAK15b}%
  \BibitemOpen
  \bibfield  {author} {\bibinfo {author} {\bibfnamefont {A.}~\bibnamefont
  {Zakharova}}, \bibinfo {author} {\bibfnamefont {M.}~\bibnamefont {Kapeller}},
  \ and\ \bibinfo {author} {\bibfnamefont {E.}~\bibnamefont {Sch{\"o}ll}},\
  }\href@noop {} {\bibfield  {journal} {\bibinfo  {journal} {J. Phys. Conf.
  Series}\ }\textbf {\bibinfo {volume} {727}},\ \bibinfo {pages} {012018}
  (\bibinfo {year} {2016})}\BibitemShut {NoStop}%
\bibitem [{\citenamefont {Hizanidis}\ \emph {et~al.}(2014)\citenamefont
  {Hizanidis}, \citenamefont {Kanas}, \citenamefont {Bezerianos},\ and\
  \citenamefont {Bountis}}]{HIZ13}%
  \BibitemOpen
  \bibfield  {author} {\bibinfo {author} {\bibfnamefont {J.}~\bibnamefont
  {Hizanidis}}, \bibinfo {author} {\bibfnamefont {V.}~\bibnamefont {Kanas}},
  \bibinfo {author} {\bibfnamefont {A.}~\bibnamefont {Bezerianos}}, \ and\
  \bibinfo {author} {\bibfnamefont {T.}~\bibnamefont {Bountis}},\ }\href
  {\doibase 10.1142/s0218127414500308} {\bibfield  {journal} {\bibinfo
  {journal} {Int. J. Bifurcation Chaos}\ }\textbf {\bibinfo {volume} {24}},\
  \bibinfo {pages} {1450030} (\bibinfo {year} {2014})}\BibitemShut {NoStop}%
\bibitem [{\citenamefont {Glaze}, \citenamefont {Lewis},\ and\ \citenamefont
  {Bahar}(2016)}]{GLA16}%
  \BibitemOpen
  \bibfield  {author} {\bibinfo {author} {\bibfnamefont {T.~A.}\ \bibnamefont
  {Glaze}}, \bibinfo {author} {\bibfnamefont {S.}~\bibnamefont {Lewis}}, \ and\
  \bibinfo {author} {\bibfnamefont {S.}~\bibnamefont {Bahar}},\ }\href@noop {}
  {\bibfield  {journal} {\bibinfo  {journal} {Chaos}\ }\textbf {\bibinfo
  {volume} {26}},\ \bibinfo {pages} {083119} (\bibinfo {year}
  {2016})}\BibitemShut {NoStop}%
\bibitem [{\citenamefont {V{\"u}llings}, \citenamefont {Sch{\"o}ll},\ and\
  \citenamefont {Lindner}(2014)}]{VUE14}%
  \BibitemOpen
  \bibfield  {author} {\bibinfo {author} {\bibfnamefont {A.}~\bibnamefont
  {V{\"u}llings}}, \bibinfo {author} {\bibfnamefont {E.}~\bibnamefont
  {Sch{\"o}ll}}, \ and\ \bibinfo {author} {\bibfnamefont {B.}~\bibnamefont
  {Lindner}},\ }\href {\doibase 10.1140/epjb/e2014-41064-y} {\bibfield
  {journal} {\bibinfo  {journal} {Eur.~Phys.~J.~B}\ }\textbf {\bibinfo {volume}
  {87}},\ \bibinfo {pages} {31} (\bibinfo {year} {2014})}\BibitemShut {NoStop}%
\bibitem [{\citenamefont {B{\"o}hm}\ \emph {et~al.}(2015)\citenamefont
  {B{\"o}hm}, \citenamefont {Zakharova}, \citenamefont {Sch{\"o}ll},\ and\
  \citenamefont {L{\"u}dge}}]{BOE15}%
  \BibitemOpen
  \bibfield  {author} {\bibinfo {author} {\bibfnamefont {F.}~\bibnamefont
  {B{\"o}hm}}, \bibinfo {author} {\bibfnamefont {A.}~\bibnamefont {Zakharova}},
  \bibinfo {author} {\bibfnamefont {E.}~\bibnamefont {Sch{\"o}ll}}, \ and\
  \bibinfo {author} {\bibfnamefont {K.}~\bibnamefont {L{\"u}dge}},\ }\href@noop
  {} {\bibfield  {journal} {\bibinfo  {journal} {Phys. Rev. E}\ }\textbf
  {\bibinfo {volume} {91}},\ \bibinfo {pages} {040901 (R)} (\bibinfo {year}
  {2015})}\BibitemShut {NoStop}%
\bibitem [{\citenamefont {Laing}(2015)}]{LAI15}%
  \BibitemOpen
  \bibfield  {author} {\bibinfo {author} {\bibfnamefont {C.~R.}\ \bibnamefont
  {Laing}},\ }\href@noop {} {\bibfield  {journal} {\bibinfo  {journal} {Phys.
  Rev. E}\ }\textbf {\bibinfo {volume} {92}},\ \bibinfo {pages} {050904(R)}
  (\bibinfo {year} {2015})}\BibitemShut {NoStop}%
\bibitem [{\citenamefont {Bera}\ and\ \citenamefont {Ghosh}(2016)}]{BER16a}%
  \BibitemOpen
  \bibfield  {author} {\bibinfo {author} {\bibfnamefont {B.~K.}\ \bibnamefont
  {Bera}}\ and\ \bibinfo {author} {\bibfnamefont {D.}~\bibnamefont {Ghosh}},\
  }\href {\doibase 10.1103/physreve.93.052223} {\bibfield  {journal} {\bibinfo
  {journal} {Phys. Rev.~E}\ }\textbf {\bibinfo {volume} {93}},\ \bibinfo
  {pages} {052223} (\bibinfo {year} {2016})}\BibitemShut {NoStop}%
\bibitem [{\citenamefont {Bogomolov}\ \emph {et~al.}(2016)\citenamefont
  {Bogomolov}, \citenamefont {Strelkova}, \citenamefont {Sch{\"o}ll},\ and\
  \citenamefont {Anishchenko}}]{BOG16a}%
  \BibitemOpen
  \bibfield  {author} {\bibinfo {author} {\bibfnamefont {S.}~\bibnamefont
  {Bogomolov}}, \bibinfo {author} {\bibfnamefont {G.}~\bibnamefont
  {Strelkova}}, \bibinfo {author} {\bibfnamefont {E.}~\bibnamefont
  {Sch{\"o}ll}}, \ and\ \bibinfo {author} {\bibfnamefont {V.~S.}\ \bibnamefont
  {Anishchenko}},\ }\href {\doibase 10.1134/s1063785016070191} {\bibfield
  {journal} {\bibinfo  {journal} {Tech. Phys. Lett.}\ }\textbf {\bibinfo
  {volume} {42}},\ \bibinfo {pages} {765} (\bibinfo {year} {2016})}\BibitemShut
  {NoStop}%
\bibitem [{\citenamefont {Zakharova}, \citenamefont {Kapeller},\ and\
  \citenamefont {Sch{\"o}ll}(2014)}]{ZAK14}%
  \BibitemOpen
  \bibfield  {author} {\bibinfo {author} {\bibfnamefont {A.}~\bibnamefont
  {Zakharova}}, \bibinfo {author} {\bibfnamefont {M.}~\bibnamefont {Kapeller}},
  \ and\ \bibinfo {author} {\bibfnamefont {E.}~\bibnamefont {Sch{\"o}ll}},\
  }\href {\doibase 10.1103/physrevlett.112.154101} {\bibfield  {journal}
  {\bibinfo  {journal} {Phys.~Rev.~Lett.}\ }\textbf {\bibinfo {volume} {112}},\
  \bibinfo {pages} {154101} (\bibinfo {year} {2014})}\BibitemShut {NoStop}%
\bibitem [{\citenamefont {Andrews}, \citenamefont {Halpern},\ and\
  \citenamefont {Purves}(1997)}]{AND97}%
  \BibitemOpen
  \bibfield  {author} {\bibinfo {author} {\bibfnamefont {T.~J.}\ \bibnamefont
  {Andrews}}, \bibinfo {author} {\bibfnamefont {S.~D.}\ \bibnamefont
  {Halpern}}, \ and\ \bibinfo {author} {\bibfnamefont {D.}~\bibnamefont
  {Purves}},\ }\href@noop {} {\bibfield  {journal} {\bibinfo  {journal} {J.
  Neurosci}\ }\textbf {\bibinfo {volume} {17}},\ \bibinfo {pages} {2859}
  (\bibinfo {year} {1997})}\BibitemShut {NoStop}%
\bibitem [{\citenamefont {Gupta}, \citenamefont {Ferguson},\ and\ \citenamefont
  {Anderson}(1998)}]{GUP98}%
  \BibitemOpen
  \bibfield  {author} {\bibinfo {author} {\bibfnamefont {S.}~\bibnamefont
  {Gupta}}, \bibinfo {author} {\bibfnamefont {N.}~\bibnamefont {Ferguson}}, \
  and\ \bibinfo {author} {\bibfnamefont {R.}~\bibnamefont {Anderson}},\
  }\href@noop {} {\bibfield  {journal} {\bibinfo  {journal} {Science}\ }\textbf
  {\bibinfo {volume} {280}},\ \bibinfo {pages} {912} (\bibinfo {year}
  {1998})}\BibitemShut {NoStop}%
\bibitem [{\citenamefont {Gog}\ and\ \citenamefont {Grenfell}(2002)}]{GOG02}%
  \BibitemOpen
  \bibfield  {author} {\bibinfo {author} {\bibfnamefont {J.~R.}\ \bibnamefont
  {Gog}}\ and\ \bibinfo {author} {\bibfnamefont {B.~T.}\ \bibnamefont
  {Grenfell}},\ }\href@noop {} {\bibfield  {journal} {\bibinfo  {journal}
  {Proc. Natl. Acad. Sci. USA}\ }\textbf {\bibinfo {volume} {99}},\ \bibinfo
  {pages} {17209} (\bibinfo {year} {2002})}\BibitemShut {NoStop}%
\bibitem [{\citenamefont {Gog}\ and\ \citenamefont {Swinton}(2002)}]{GOG02a}%
  \BibitemOpen
  \bibfield  {author} {\bibinfo {author} {\bibfnamefont {J.~R.}\ \bibnamefont
  {Gog}}\ and\ \bibinfo {author} {\bibfnamefont {J.}~\bibnamefont {Swinton}},\
  }\href@noop {} {\bibfield  {journal} {\bibinfo  {journal} {J. Math. Biol.}\
  }\textbf {\bibinfo {volume} {44}},\ \bibinfo {pages} {169} (\bibinfo {year}
  {2002})}\BibitemShut {NoStop}%
\bibitem [{\citenamefont {Gomes}, \citenamefont {Medley},\ and\ \citenamefont
  {Nokes}(2002)}]{GOM02}%
  \BibitemOpen
  \bibfield  {author} {\bibinfo {author} {\bibfnamefont {M.~G.~M.}\
  \bibnamefont {Gomes}}, \bibinfo {author} {\bibfnamefont {G.~F.}\ \bibnamefont
  {Medley}}, \ and\ \bibinfo {author} {\bibfnamefont {D.~J.}\ \bibnamefont
  {Nokes}},\ }\href@noop {} {\bibfield  {journal} {\bibinfo  {journal} {Proc.
  R. Soc. Lond. [Biol.]}\ }\textbf {\bibinfo {volume} {269}},\ \bibinfo {pages}
  {227} (\bibinfo {year} {2002})}\BibitemShut {NoStop}%
\bibitem [{\citenamefont {Koelle}\ \emph {et~al.}(2006)\citenamefont {Koelle},
  \citenamefont {Cobey}, \citenamefont {Grenfell},\ and\ \citenamefont
  {Pascual}}]{KOE06}%
  \BibitemOpen
  \bibfield  {author} {\bibinfo {author} {\bibfnamefont {K.}~\bibnamefont
  {Koelle}}, \bibinfo {author} {\bibfnamefont {S.}~\bibnamefont {Cobey}},
  \bibinfo {author} {\bibfnamefont {B.~T.}\ \bibnamefont {Grenfell}}, \ and\
  \bibinfo {author} {\bibfnamefont {M.}~\bibnamefont {Pascual}},\ }\href
  {\doibase 10.1126/science.1132745} {\bibfield  {journal} {\bibinfo  {journal}
  {Science}\ }\textbf {\bibinfo {volume} {314}},\ \bibinfo {pages} {1898}
  (\bibinfo {year} {2006})},\ \Eprint
  {http://arxiv.org/abs/http://www.sciencemag.org/content/314/5807/1898.full.pdf}
  {http://www.sciencemag.org/content/314/5807/1898.full.pdf} \BibitemShut
  {NoStop}%
\bibitem [{\citenamefont {Koelle}, \citenamefont {Kamradt},\ and\ \citenamefont
  {Pascual}(2009)}]{KOE09}%
  \BibitemOpen
  \bibfield  {author} {\bibinfo {author} {\bibfnamefont {K.}~\bibnamefont
  {Koelle}}, \bibinfo {author} {\bibfnamefont {M.}~\bibnamefont {Kamradt}}, \
  and\ \bibinfo {author} {\bibfnamefont {M.}~\bibnamefont {Pascual}},\ }\href
  {\doibase 10.1016/j.epidem.2009.05.003} {\bibfield  {journal} {\bibinfo
  {journal} {Epidemics}\ }\textbf {\bibinfo {volume} {1}},\ \bibinfo {pages}
  {129} (\bibinfo {year} {2009})}\BibitemShut {NoStop}%
\bibitem [{\citenamefont {Calvez}, \citenamefont {Korobeinikov},\ and\
  \citenamefont {Maini}(2005)}]{CAL05}%
  \BibitemOpen
  \bibfield  {author} {\bibinfo {author} {\bibfnamefont {V.}~\bibnamefont
  {Calvez}}, \bibinfo {author} {\bibfnamefont {A.}~\bibnamefont
  {Korobeinikov}}, \ and\ \bibinfo {author} {\bibfnamefont {P.}~\bibnamefont
  {Maini}},\ }\href@noop {} {\bibfield  {journal} {\bibinfo  {journal}
  {J.~Theor. Biol.}\ }\textbf {\bibinfo {volume} {233}},\ \bibinfo {pages} {75}
  (\bibinfo {year} {2005})}\BibitemShut {NoStop}%
\bibitem [{\citenamefont {Adams}\ and\ \citenamefont {Sasaki}(2007)}]{ADA07}%
  \BibitemOpen
  \bibfield  {author} {\bibinfo {author} {\bibfnamefont {B.}~\bibnamefont
  {Adams}}\ and\ \bibinfo {author} {\bibfnamefont {A.}~\bibnamefont {Sasaki}},\
  }\href@noop {} {\bibfield  {journal} {\bibinfo  {journal} {Math. Biosci.}\
  }\textbf {\bibinfo {volume} {210}},\ \bibinfo {pages} {680} (\bibinfo {year}
  {2007})}\BibitemShut {NoStop}%
\bibitem [{\citenamefont {Minayev}\ and\ \citenamefont
  {Ferguson}(2008)}]{MIN08a}%
  \BibitemOpen
  \bibfield  {author} {\bibinfo {author} {\bibfnamefont {P.}~\bibnamefont
  {Minayev}}\ and\ \bibinfo {author} {\bibfnamefont {N.}~\bibnamefont
  {Ferguson}},\ }\href {\doibase 10.1098/rsif.2008.0333} {\bibfield  {journal}
  {\bibinfo  {journal} {J. R. Soc. Interface}\ }\textbf {\bibinfo {volume}
  {6}},\ \bibinfo {pages} {509} (\bibinfo {year} {2008})}\BibitemShut {NoStop}%
\bibitem [{\citenamefont {Cobey}\ and\ \citenamefont {Pascual}(2011)}]{COB11}%
  \BibitemOpen
  \bibfield  {author} {\bibinfo {author} {\bibfnamefont {S.}~\bibnamefont
  {Cobey}}\ and\ \bibinfo {author} {\bibfnamefont {M.}~\bibnamefont
  {Pascual}},\ }\href@noop {} {\bibfield  {journal} {\bibinfo  {journal}
  {Journal of theoretical biology}\ }\textbf {\bibinfo {volume} {270}},\
  \bibinfo {pages} {80} (\bibinfo {year} {2011})}\BibitemShut {NoStop}%
\bibitem [{\citenamefont {Recker}\ \emph {et~al.}(2009)\citenamefont {Recker},
  \citenamefont {Blyuss}, \citenamefont {Simmons}, \citenamefont {Hien},
  \citenamefont {Wills}, \citenamefont {Farrar},\ and\ \citenamefont
  {Gupta}}]{REC09}%
  \BibitemOpen
  \bibfield  {author} {\bibinfo {author} {\bibfnamefont {M.}~\bibnamefont
  {Recker}}, \bibinfo {author} {\bibfnamefont {K.~B.}\ \bibnamefont {Blyuss}},
  \bibinfo {author} {\bibfnamefont {C.~P.}\ \bibnamefont {Simmons}}, \bibinfo
  {author} {\bibfnamefont {T.~T.}\ \bibnamefont {Hien}}, \bibinfo {author}
  {\bibfnamefont {B.}~\bibnamefont {Wills}}, \bibinfo {author} {\bibfnamefont
  {J.}~\bibnamefont {Farrar}}, \ and\ \bibinfo {author} {\bibfnamefont
  {S.}~\bibnamefont {Gupta}},\ }\href@noop {} {\bibfield  {journal} {\bibinfo
  {journal} {Proc. R.~Soc.~B}\ }\textbf {\bibinfo {volume} {276}},\ \bibinfo
  {pages} {2541} (\bibinfo {year} {2009})}\BibitemShut {NoStop}%
\bibitem [{\citenamefont {Kucharski}, \citenamefont {Andreasen},\ and\
  \citenamefont {Gog}(2016)}]{KUC16}%
  \BibitemOpen
  \bibfield  {author} {\bibinfo {author} {\bibfnamefont {A.~J.}\ \bibnamefont
  {Kucharski}}, \bibinfo {author} {\bibfnamefont {V.}~\bibnamefont
  {Andreasen}}, \ and\ \bibinfo {author} {\bibfnamefont {J.~R.}\ \bibnamefont
  {Gog}},\ }\href@noop {} {\bibfield  {journal} {\bibinfo  {journal} {Journal
  of mathematical biology}\ }\textbf {\bibinfo {volume} {72}},\ \bibinfo
  {pages} {1} (\bibinfo {year} {2016})}\BibitemShut {NoStop}%
\bibitem [{\citenamefont {Blyuss}\ and\ \citenamefont
  {Kyrychko}(2012)}]{BLY12}%
  \BibitemOpen
  \bibfield  {author} {\bibinfo {author} {\bibfnamefont {K.~B.}\ \bibnamefont
  {Blyuss}}\ and\ \bibinfo {author} {\bibfnamefont {Y.~N.}\ \bibnamefont
  {Kyrychko}},\ }\href {\doibase 10.1007/s11538-012-9763-8} {\bibfield
  {journal} {\bibinfo  {journal} {Bull. Math. Biol.}\ }\textbf {\bibinfo
  {volume} {74}},\ \bibinfo {pages} {2488} (\bibinfo {year}
  {2012})}\BibitemShut {NoStop}%
\bibitem [{\citenamefont {Blyuss}(2013)}]{BLY13}%
  \BibitemOpen
  \bibfield  {author} {\bibinfo {author} {\bibfnamefont {K.~B.}\ \bibnamefont
  {Blyuss}},\ }\href@noop {} {\bibfield  {journal} {\bibinfo  {journal} {J.
  Math. Biol.}\ }\textbf {\bibinfo {volume} {66}},\ \bibinfo {pages} {115}
  (\bibinfo {year} {2013})}\BibitemShut {NoStop}%
\bibitem [{\citenamefont {Blyuss}(2014)}]{BLY14}%
  \BibitemOpen
  \bibfield  {author} {\bibinfo {author} {\bibfnamefont {K.~B.}\ \bibnamefont
  {Blyuss}},\ }\href@noop {} {\bibfield  {journal} {\bibinfo  {journal} {J.
  Math. Biol}\ }\textbf {\bibinfo {volume} {69}},\ \bibinfo {pages} {1431}
  (\bibinfo {year} {2014})}\BibitemShut {NoStop}%
\bibitem [{\citenamefont {Charles}\ and\ \citenamefont {Baca}(2013)}]{CHA13a}%
  \BibitemOpen
  \bibfield  {author} {\bibinfo {author} {\bibfnamefont {A.~C.}\ \bibnamefont
  {Charles}}\ and\ \bibinfo {author} {\bibfnamefont {S.~M.}\ \bibnamefont
  {Baca}},\ }\href@noop {} {\bibfield  {journal} {\bibinfo  {journal} {Nat.
  Rev. Neurol.}\ } (\bibinfo {year} {2013})}\BibitemShut {NoStop}%
\bibitem [{\citenamefont {Chapman}\ and\ \citenamefont
  {Mesbahi}(2013)}]{CHA13b}%
  \BibitemOpen
  \bibfield  {author} {\bibinfo {author} {\bibfnamefont {A.}~\bibnamefont
  {Chapman}}\ and\ \bibinfo {author} {\bibfnamefont {M.}~\bibnamefont
  {Mesbahi}},\ }in\ \href {\doibase 10.1109/acc.2013.6580798} {\emph {\bibinfo
  {booktitle} {American Control Conference (ACC), 2013}}}\ (\bibinfo {year}
  {2013})\ pp.\ \bibinfo {pages} {6126--6131}\BibitemShut {NoStop}%
\bibitem [{\citenamefont {Kyrychko}\ and\ \citenamefont
  {Blyuss}(2005)}]{KYR05}%
  \BibitemOpen
  \bibfield  {author} {\bibinfo {author} {\bibfnamefont {Y.~N.}\ \bibnamefont
  {Kyrychko}}\ and\ \bibinfo {author} {\bibfnamefont {K.~B.}\ \bibnamefont
  {Blyuss}},\ }\href@noop {} {\bibfield  {journal} {\bibinfo  {journal}
  {Nonlin. Anal. RWA}\ }\textbf {\bibinfo {volume} {6}},\ \bibinfo {pages}
  {495} (\bibinfo {year} {2005})}\BibitemShut {NoStop}%
\bibitem [{\citenamefont {Blyuss}\ and\ \citenamefont
  {Kyrychko}(2010)}]{BLY10}%
  \BibitemOpen
  \bibfield  {author} {\bibinfo {author} {\bibfnamefont {K.~B.}\ \bibnamefont
  {Blyuss}}\ and\ \bibinfo {author} {\bibfnamefont {Y.~N.}\ \bibnamefont
  {Kyrychko}},\ }\href@noop {} {\bibfield  {journal} {\bibinfo  {journal}
  {Bull. Math. Biol.}\ }\textbf {\bibinfo {volume} {72}},\ \bibinfo {pages}
  {490} (\bibinfo {year} {2010})}\BibitemShut {NoStop}%
\bibitem [{\citenamefont {Engelborghs}, \citenamefont {Luzyanina},\ and\
  \citenamefont {Samaey}(2001)}]{ENG01}%
  \BibitemOpen
  \bibfield  {author} {\bibinfo {author} {\bibfnamefont {K.}~\bibnamefont
  {Engelborghs}}, \bibinfo {author} {\bibfnamefont {T.}~\bibnamefont
  {Luzyanina}}, \ and\ \bibinfo {author} {\bibfnamefont {G.}~\bibnamefont
  {Samaey}},\ }\href@noop {} {\enquote {\bibinfo {title} {{DDE}-{BIFTOOL} v.
  2.00: a matlab package for bifurcation analysis of delay differential
  equations},}\ }\bibinfo {type} {Tech. Rep.}\ \bibinfo {number} {TW-330}\
  (\bibinfo  {institution} {Department of Computer Science, K.U.Leuven},\
  \bibinfo {address} {Belgium},\ \bibinfo {year} {2001})\BibitemShut {NoStop}%
\bibitem [{\citenamefont {Blyuss}\ and\ \citenamefont {Gupta}(2009)}]{BLY09}%
  \BibitemOpen
  \bibfield  {author} {\bibinfo {author} {\bibfnamefont {K.~B.}\ \bibnamefont
  {Blyuss}}\ and\ \bibinfo {author} {\bibfnamefont {S.}~\bibnamefont {Gupta}},\
  }\href@noop {} {\bibfield  {journal} {\bibinfo  {journal} {J. Math. Biol.}\
  }\textbf {\bibinfo {volume} {58}},\ \bibinfo {pages} {923} (\bibinfo {year}
  {2009})}\BibitemShut {NoStop}%
\bibitem [{\citenamefont {Shampine}\ and\ \citenamefont
  {Thompson}(2001)}]{SHA01a}%
  \BibitemOpen
  \bibfield  {author} {\bibinfo {author} {\bibfnamefont {L.~F.}\ \bibnamefont
  {Shampine}}\ and\ \bibinfo {author} {\bibfnamefont {S.}~\bibnamefont
  {Thompson}},\ }\href@noop {} {\bibfield  {journal} {\bibinfo  {journal}
  {Appl. Num. Math.}\ }\textbf {\bibinfo {volume} {37}},\ \bibinfo {pages}
  {441} (\bibinfo {year} {2001})}\BibitemShut {NoStop}%
\bibitem [{\citenamefont {Zakharova}\ \emph {et~al.}(2013)\citenamefont
  {Zakharova}, \citenamefont {Feoktistov}, \citenamefont {Vadivasova},\ and\
  \citenamefont {Sch{\"o}ll}}]{ZAK13}%
  \BibitemOpen
  \bibfield  {author} {\bibinfo {author} {\bibfnamefont {A.}~\bibnamefont
  {Zakharova}}, \bibinfo {author} {\bibfnamefont {A.}~\bibnamefont
  {Feoktistov}}, \bibinfo {author} {\bibfnamefont {T.}~\bibnamefont
  {Vadivasova}}, \ and\ \bibinfo {author} {\bibfnamefont {E.}~\bibnamefont
  {Sch{\"o}ll}},\ }\href {\doibase 10.1140/epjst/e2013-02031-x} {\bibfield
  {journal} {\bibinfo  {journal} {Eur. Phys. J. Spec. Top.}\ }\textbf {\bibinfo
  {volume} {222}},\ \bibinfo {pages} {2481} (\bibinfo {year}
  {2013})}\BibitemShut {NoStop}%
\bibitem [{\citenamefont {Omel'chenko}, \citenamefont {Wolfrum},\ and\
  \citenamefont {Maistrenko}(2010)}]{OME10a}%
  \BibitemOpen
  \bibfield  {author} {\bibinfo {author} {\bibfnamefont {O.~E.}\ \bibnamefont
  {Omel'chenko}}, \bibinfo {author} {\bibfnamefont {M.}~\bibnamefont
  {Wolfrum}}, \ and\ \bibinfo {author} {\bibfnamefont {Y.}~\bibnamefont
  {Maistrenko}},\ }\href {\doibase 10.1103/physreve.81.065201} {\bibfield
  {journal} {\bibinfo  {journal} {Phys. Rev.~E}\ }\textbf {\bibinfo {volume}
  {81}},\ \bibinfo {pages} {065201(R)} (\bibinfo {year} {2010})}\BibitemShut
  {NoStop}%
\bibitem [{\citenamefont {Ballesteros}, \citenamefont {Vergu},\ and\
  \citenamefont {Cazelles}(2009)}]{BAL09a}%
  \BibitemOpen
  \bibfield  {author} {\bibinfo {author} {\bibfnamefont {S.}~\bibnamefont
  {Ballesteros}}, \bibinfo {author} {\bibfnamefont {E.}~\bibnamefont {Vergu}},
  \ and\ \bibinfo {author} {\bibfnamefont {B.}~\bibnamefont {Cazelles}},\
  }\href@noop {} {\bibfield  {journal} {\bibinfo  {journal} {PLoS One}\
  }\textbf {\bibinfo {volume} {4}},\ \bibinfo {pages} {e7426} (\bibinfo {year}
  {2009})}\BibitemShut {NoStop}%
\bibitem [{\citenamefont {Omelchenko}\ \emph {et~al.}(2011)\citenamefont
  {Omelchenko}, \citenamefont {Maistrenko}, \citenamefont {H{\"o}vel},\ and\
  \citenamefont {Sch{\"o}ll}}]{OME11}%
  \BibitemOpen
  \bibfield  {author} {\bibinfo {author} {\bibfnamefont {I.}~\bibnamefont
  {Omelchenko}}, \bibinfo {author} {\bibfnamefont {Y.}~\bibnamefont
  {Maistrenko}}, \bibinfo {author} {\bibfnamefont {P.}~\bibnamefont
  {H{\"o}vel}}, \ and\ \bibinfo {author} {\bibfnamefont {E.}~\bibnamefont
  {Sch{\"o}ll}},\ }\href {\doibase 10.1103/physrevlett.106.234102} {\bibfield
  {journal} {\bibinfo  {journal} {Phys. Rev. Lett.}\ }\textbf {\bibinfo
  {volume} {106}},\ \bibinfo {pages} {234102} (\bibinfo {year}
  {2011})}\BibitemShut {NoStop}%
\bibitem [{\citenamefont {Omelchenko}\ \emph {et~al.}(2012)\citenamefont
  {Omelchenko}, \citenamefont {Riemenschneider}, \citenamefont {H{\"o}vel},
  \citenamefont {Maistrenko},\ and\ \citenamefont {Sch{\"o}ll}}]{OME12}%
  \BibitemOpen
  \bibfield  {author} {\bibinfo {author} {\bibfnamefont {I.}~\bibnamefont
  {Omelchenko}}, \bibinfo {author} {\bibfnamefont {B.}~\bibnamefont
  {Riemenschneider}}, \bibinfo {author} {\bibfnamefont {P.}~\bibnamefont
  {H{\"o}vel}}, \bibinfo {author} {\bibfnamefont {Y.}~\bibnamefont
  {Maistrenko}}, \ and\ \bibinfo {author} {\bibfnamefont {E.}~\bibnamefont
  {Sch{\"o}ll}},\ }\href {\doibase 10.1103/physreve.85.026212} {\bibfield
  {journal} {\bibinfo  {journal} {Phys. Rev.~E}\ }\textbf {\bibinfo {volume}
  {85}},\ \bibinfo {pages} {026212} (\bibinfo {year} {2012})}\BibitemShut
  {NoStop}%
\end{thebibliography}


%

\end{document}